\documentclass[12pt,leqno]{article}
\title{Gravitational instability of finite isothermal spheres}
\author{Pierre-Henri Chavanis$^{1,2}$}
\date{}

\usepackage{amsthm,amsmath,amssymb,a4wide,psfig}
\def\mb#1{\setbox0=\hbox{$#1$}\kern-.025em\copy0\kern-\wd0
\kern-0.05em\copy0\kern-\wd0\kern-.025em\raise.0233em\box0}

\begin{document}
\maketitle
\vspace*{-1cm}
\begin{center}
$^{1}$ Laboratoire de Physique Quantique,
Universit\'e Paul Sabatier,\\
118 route de Narbonne 31062 Toulouse, France.\\

$^{2}$ Institute for Theoretical Physics,
University of California, Santa Barbara, California.\\
{\footnotesize
e-mail: chavanis@irsamc2.ups-tlse.fr; web: http://w3-phystheo.ups-tlse.fr/chavanis/}\\

\vspace{0.5cm}
\end{center}

\begin{abstract}

We investigate the stability of bounded self-gravitating systems in
the canonical ensemble by using a thermodynamical approach. Our study
extends the earlier work of Padmanabhan [ApJ Supp. {71}, 651
(1989)] in the microcanonical ensemble. By studying the second
variations of the free energy, we find that instability sets in
precisely at the point of minimum temperature in agreement with the
theorem of Katz [MNRAS {183}, 765 (1978)]. The perturbation that
induces instability at this point is calculated explicitly; it has
{\it not} a ``core-halo'' structure contrary to what happens in the
microcanonical ensemble. We also study Jeans type gravitational
instability of isothermal gaseous spheres described by Navier-Stokes
equations.  The introduction of a container and the consideration of
an inhomogeneous distribution of matter avoids the Jeans swindle. We show
analytically the equivalence between dynamical stability and
thermodynamical stability and the fact that the stability of
isothermal gas spheres does not depend on the viscosity. This confirms
the findings of Semelin {\it et al.}  [Phys. Rev. D {63} 084005
(2001)] who used numerical methods or approximations. We also give a
simpler derivation of the geometric hierarchy of scales inducing
instability discovered by these authors. The density profiles that
trigger these instabilities are calculated analytically; high order
modes of instability present numerous oscillations that also follow a
geometric progression. This suggests that the system will fragment in
a series of `clumps' and that these `clumps' will themselves fragment
in substructures. The fact that both the domain sizes leading to
instability and the `clumps' sizes within a domain follow a geometric
progression with the same ratio suggests a fractal-like behavior. This
gives further support to the interpretation of de Vega et al. [Nature,
383, 56 (1996)].

\end{abstract}

\section{Introduction}
\label{sec_introduction}

The thermodynamics of self-gravitating systems is a fascinating
subject. It started with Antonov (1962)'s discovery that, when a
self-gravitating system is confined within a box of radius $R$, no
maximum entropy state can exist below a certain critical energy
$E=-0.335GM^{2}/R$. This intriguing result was further discussed by
Lynden-Bell \& Wood (1968) who conjectured that for $E<-0.335GM^{2}/R$
the system would collapse and overheat.  This is called ``gravothermal
catastrophe'' or ``Antonov instability''. Lynden-Bell \& Wood 
have related this phenomenon to the very particular property of
self-gravitating systems to possess {\it negative} specific heats. The
gravothermal catastrophe picture has been confirmed by sophisticated
numerical simulations (Larson 1970, Cohn 1980, Lynden-Bell \& Eggleton
1980) and is expected to play a crucial role in the evolution of
globular clusters. It is found that the collapse proceeds
self-similarly (with power law behaviors) and that the central
density becomes infinite in a finite time. This instability has been
known as ``core collapse'' and many globular clusters have probably
experienced core collapse (Binney \& Tremaine 1987). In the case of
dense clusters of compact stars (neutron stars or stellar mass black
holes), the gravothermal catastrophe can lead to the formation of
supermassive black holes of the right size to explain quasars and AGNs
(Shapiro \& Teukolsky 1995). Statistical mechanics is also relevant
for collisionless stellar systems (e.g., elliptical galaxies, dark
matter,...) undergoing a violent relaxation (Lynden-Bell 1967,
Chavanis {\it et al.} 1996, Chavanis \& Sommeria 1998, Chavanis
1998a,2001a). In particular, the inner regions of elliptical galaxies
are close to isothermal and this is an important ingredient to
understand de Vaucouleurs' $R^{1/4}$ law (Hjorth \& Madsen 1993).

On a theoretical point of view, the stability of isothermal spheres
has been first investigated by Katz (1978) with a very powerful method
extending Poincar\'e's theory of linear series of equilibrium.  He
found that instability sets in precisely at the point of minimum
energy. This stability analysis was reconsidered by Padmanabhan (1989)
who studied the sign of the second variations of entropy and reduced
the problem of stability to an eigenvalue equation. This leads to the
same stability limit as Katz but the method of Padmanabhan provides in
addition the form of the perturbation that induces instability at the
critical point. It is found that this perturbation presents a
``core-halo'' structure.

The analysis of Padmanabhan (1989) was performed in the microcanonical
ensemble in which the energy is fixed. The microcanonical ensemble is
probably the most relevant for studying stellar systems like
elliptical galaxies or globular clusters (Binney \& Tremaine
1987). Indeed, apart from a slow evaporation, these systems can be
assumed isolated so the evolution conserves energy $E$ and mass
$M$. In addition, from the viewpoint of statistical mechanics, only
the microcanonical ensemble is rigorously justified for non extensive
systems, as discussed in the review of Padmanabhan (1990). However, it
is always possible to define formally a canonical ensemble or a grand
canonical ensemble in which the temperature is fixed instead of the
energy. As suggested by de Vega {\it et al.} (1996a, 1996b) these
ensembles may be suitable for describing the cold interstellar medium
where the temperature is imposed by the cosmic background radiation at
$T\sim 3K$ in the outer parts of galaxies, devoid of any star and
heating sources (Pfenninger {\it et al.} 1994, Pfenninger \& Combes
1994).  In particular, by working out the statistical mechanics of the
self-gravitating gas, de Vega {\it et al.} (1996a, 1996b) have shown
that self-gravity can provide a dynamical mechanism to produce the
fractal structure of the interstellar medium.  They used the same
approach to explain the fractal structure of the universe (de Vega
{\it et al.} 1998), assuming that galaxies have reached a
quasi-thermodynamical equilibrium like in the early work of Saslaw \&
Hamilton (1984).

For self-gravitating systems, it is well known that the
thermodynamical ensembles do not coincide in the whole range of
parameters (Padmanabhan 1990). Using toy models, Lynden-Bell \&
Lynden-Bell (1977) and Padmanabhan (1990) demonstrated that the region
of negative specific heats allowed in the microcanonical ensemble is
replaced by a phase transition in the canonical ensemble. This phase
transition separates a dilute ``gaseous'' phase from a dense
``collapsed'' phase. Since these toy models are not very realistic,
the self-gravitating gas was also studied in a meanfield
approximation. In this viewpoint, an isothermal sphere is stable if and
only if it is a local maximum of an appropriate thermodynamical
potential (the entropy in the microcanonical ensemble and the free
energy in the canonical ensemble). As expected from physical grounds,
phase transition occurs when the gaseous sphere ceases to be a local
maximum of this potential and becomes a saddle point.  In this paper,
we investigate the stability of isothermal gaseous spheres in the
canonical ensemble by studying the sign of the second variations of
the free energy. Our analysis is a direct extension of Padmanabhan
(1989)'s approach in the microcanonical ensemble. The two studies
therefore provide a unified description of the stability of isothermal
spheres in the meanfield approximation in terms of thermodynamical
potentials.

For a long time, the thermodynamics of self-gravitating systems was
considered exclusively in a meanfield approach (or with toy
models). However, recently, de Vega \& Sanchez (2001a,b) have
developed a rigorous statistical mechanics of self-gravitating
systems by using field theoretical methods. In particular, they
evidenced the existence of a thermodynamic limit in which the number
of particles $N$ and the volume $R^{3}$ go to infinity keeping $N/R$
fixed. This very unusual thermodynamic limit proves to be appropriate
to non extensive systems. By using Monte Carlo simulations and
analytical calculations, they showed that the meanfield approximation
correctly describes the thermodynamic limit except near the critical
points where a phase transition occurs. They also {\it derived} the
equation of state of a self-gravitating gas instead of assuming it, as
is done in the meanfield treatments. Therefore, their work fully
justifies the studies of previous authors and specifies their range of
validity. 

This paper is organized as follows. In section \ref{sec_thermo}, we
introduce a meanfield description of the system in the canonical
ensemble and show that critical points of free energy $J$ at fixed
temperature $T$ and mass $M$ correspond to isothermal spheres like
those studied in the context of stellar structure (Chandraskhar
1942). We show that there is no global
maximum of free energy.  There is not even a local maximum of free
energy in an unbounded domain: unbounded isothermal spheres have an
infinite mass! We restrict therefore our analysis
to the case of self-gravitating systems confined within a spherical
box of radius $R$. In this case, there exists {\it local} maxima of
$J$ (metastable states) if the normalized temperature $\eta={\beta
GM\over R}$ is less than $2.52$ and the density contrast ${\cal
R}=\rho(0)/\rho(R)<32.1$. Critical points of free energy with density
contrast ${\cal R}>32.1$ are unstable saddle points. For
$\eta>2.52$, there are not even critical points of free energy: in
that case, the system is expected to undergo a phase transition and
collapse. This ``isothermal collapse'' is the counterpart of the
``gravothermal catastrophe'' in the microcanonical 
ensemble (see Figs. \ref{gravotherme}-\ref{isocollapse}).

In section \ref{sec_pad}, we study the sign of the second variations
of free energy by using the methods of Padmanabhan (1989) introduced
in the microcanonical ensemble. We show analytically that instability
sets in precisely at the point of minimum temperature in agreement
with the theorem of Katz (1978). The perturbation that induces
instability at this point is calculated explicitly; it has {\it not} a
``core-halo'' structure contrary to what happens in the microcanonical
ensemble.

In section \ref{sec_dyn}, we study Jeans type gravitational
instability of isothermal gaseous spheres described by Navier-Stokes
equations. The introduction of a container removes the problems
associated with an infinite homogeneous medium and avoids the Jeans
swindle. We show analytically the equivalence between dynamical
stability and thermodynamical stability and the fact that the
stability of isothermal gas spheres does not depend on the
viscosity. This confirms the findings of Semelin {\it et al.} (2001)
who used numerical methods or approximations. We also give a simpler
derivation of the geometric hierarchy of scales inducing instability
discovered by these authors using sophisticated renormalization group
technics (Semelin {\it et al.} 1999). This provides a more
illuminating interpretation of their results. 

In section \ref{sec_frag}, we make speculations about the
fragmentation and the fractal structure of an isothermal
self-gravitating gas.  We distinguish between the Jeans length $L_{J}$
defined with the mean density and the King's length $L_{K}$ defined
with the central density. If we fix the Jeans length, instability
occurs for $R>\sqrt{2.52...\over 3}L_{J}$ and is marked by the {\it
absence} of critical point of free energy (i.e., hydrostatic
equilibrium) above this threshhold. In that case the system is
expected to collapse without fragmenting. If we fix the King's length
(or core size), a first instability occurs for $R={8.99...\over
3}L_{K}$. Above this threshold critical points of free energy {\it
still} exist but they are unstable saddle points. Secondary
instabilities occur at larger box radii that asymptotically follow a
geometric progression $R_{n}\sim [10.74...]^{n} L_{K}$.  The density
profiles that trigger these high order modes of instability are
calculated analytically. They present more and more oscillations whose
zeros also follow a geometric progression $r_{n}\sim
[10.74...]^{n}R$. The profile that destabilizes the singular
isothermal sphere has an infinite number of nodes! Each oscillation
can be interpreted as a `germ' in the langage of phase transition
and the above picture suggests that the system will fragment into a
series of `clumps'.  It is expected that these `clumps' will evolve
by achieving higher and higher density contrasts, and finally fragment
in turn into substructures.  This yields a hierarchy of structures
fitting one into each other in a self-similar way. This picture is
given further support by the fact that {\it both} the domain sizes
inducing instability and the zeros of the perturbation profile in each
domain follow a geometric progression with the same ratio. This
double-geometric progression may explain in a natural way the fractal
structure of a self-gravitating gas like the interstellar medium and
the large scale structures of the universe. This gives further support
to the interpretation of de Vega {\it et al.} (1996a, 1996b, 1998) who
first realized the importance played by self-gravity in building a
fractal distribution of matter.

\section{Thermodynamical stability of self-gravitating systems in the canonical ensemble}
\label{sec_thermo}

\subsection{The free energy}
\label{sec_massieu}

Consider a system of $N$ particles, each of mass $m$, interacting via
Newtonian gravity. The particles can be galaxies, stars, atoms,
etc... We assume that the system is non rotating and non
expanding. Let $f({\bf r},{\bf v},t)$ denote the distribution function
of the system, i.e. $f({\bf r},{\bf v},t)d^{3}{\bf r}d^{3}{\bf v}$
gives the mass of particles whose position and velocity are in the
cell $({\bf r},{\bf v};{\bf r}+d^{3}{\bf r},{\bf v}+d^{3}{\bf v})$ at
time $t$. The integral of $f$ over the velocity determines the spatial
density
\begin{equation}
\rho=\int f d^{3}{\bf v}.
\label{rho}
\end{equation} 
On the other hand, in the meanfield approximation, the total mass and
the total energy of the system can be expressed as
\begin{equation}
M=Nm=\int \rho d^{3}{\bf r},
\label{M}
\end{equation} 
\begin{equation}
E={1\over 2}\int f v^{2} d^{3}{\bf r}d^{3}{\bf v}+{1\over 2}\int \rho \Phi d^{3}{\bf r} =K+W,
\label{E}
\end{equation} 
where $K$ is the kinetic energy and $W$ the potential energy. The gravitational potential $\Phi$ is related to the star density by the Newton-Poisson equation
\begin{equation}
\Delta \Phi =4\pi G \rho.
\label{Poisson}
\end{equation} 
Finally, the Boltzmann entropy is given by the standard formula
\begin{equation}
S=-k\int {f\over m} \ln {f\over m}  d^{3}{\bf r}d^{3}{\bf v},
\label{Bentropy}
\end{equation} 
which can be obtained by counting the number of microstates corresponding to a given macrostate and taking the logarithm of this number (Ogorodnikov 1965).

We shall work in the canonical ensemble in which the temperature is fixed, allowing the energy to fluctuate. In that case, the relevant thermodynamical potential is the Massieu function related to the Helmholtz free energy $F=E-TS$ by $J=-{1\over T}F$. Hence
\begin{equation}
J=S-{1\over T}E.
\label{J}
\end{equation} 
At equilibrium, the system is expected to be in a state that maximizes the Massieu function (\ref{J}) for a fixed total mass $M$ (in the following we shall call $J$ the free energy, although this is only the free energy up to a negative proportionality factor). 

\subsection{The isothermal gaseous spheres}
\label{sec_isothermal}

Following Padmanabhan (1989)'s procedure, we start to maximize the
free energy $J$ for a given density field $\rho({\bf r})$. This yields
the Maxwell-Boltzmann distribution
\begin{equation}
f=\biggl ({m\over 2\pi k T}\biggr )^{3/2}\rho({\bf r})e^{-{mv^{2}\over 2kT}},
\label{MB}
\end{equation} 
which is a {\it global} maximum of $J$ with the previous
constraint. Substituting this optimal distribution function in
Eqs. (\ref{E})-(\ref{Bentropy}), we can express the energy and the
entropy in terms of the spatial density in the form
\begin{equation}
E={3\over 2}NkT+{1\over 2}\int \rho\Phi d^{3}{\bf r},
\label{Erho}
\end{equation}
\begin{equation}
{S\over k}={3N\over 2}+{3N\over 2}\ln \biggl ({2\pi kT\over m}\biggr )-\int{\rho\over m}\ln {\rho\over m}d^{3}{\bf r}.
\label{Srho}
\end{equation}
We can now determine the variations of $J$ around a given density profile $\rho({\bf r})$. To second order in the expansion we get
\begin{eqnarray}
\delta J=-{k\over m}\int \delta\rho \biggl (\ln {\rho\over m}+1\biggr )d^{3}{\bf r}-{k\over m}\int {(\delta\rho)^{2}\over 2\rho} d^{3}{\bf r}\nonumber\\
-{1\over T}\int \Phi\delta\rho d^{3}{\bf r}-{1\over 2T}\int \delta\rho\delta\Phi d^{3}{\bf r}.
\label{d2J}
\end{eqnarray}

Introducing a Lagrange multiplier $\alpha$ to satisfy the conservation of mass, the condition that $J$ is an extremum is written (to first order)
\begin{eqnarray}
0=\delta J-\alpha\delta M=-\int d^{3}{\bf r}\biggl \lbrack {\Phi\over T}+{k\over m}\biggl (\ln {\rho\over m}+1\biggr )+\alpha\biggr \rbrack \delta\rho. 
\label{dJ1}
\end{eqnarray}  
This condition must be satisfied for any variations $\delta\rho$. This yields the Boltzmann distribution
\begin{equation}
\rho =A e^{-\beta \Phi},
\label{rhoB}
\end{equation}
where we have set
\begin{equation}
\beta={m\over kT}.
\label{beta}
\end{equation}
Therefore, the Boltzmann distribution (\ref{rhoB}) is a {\it critical}
point of free energy. This does not insure, however, that it is a {\it
maximum} of $J$. It is not even clear that the extremum problem
leading to Eq. (\ref{rhoB}) has a solution. Indeed, the gravitational
potential that appears in Eq. (\ref{rhoB}) must be determined
self-consistently by solving the mean field equation
\begin{equation}
\Delta \Phi =4\pi G A e^{-\beta \Phi}, 
\label{Mf}
\end{equation} 
obtained by substituting the density (\ref{rhoB}) in the Poisson equation (\ref{Poisson}), and relating the Lagrange multiplier $A$ to the constraint $M$. As we shall see, this problem does not always have a solution. When a solution exists, we must consider the sign of the second order variations $\delta^{2}J$ to determine whether it is a maximum or not.

\subsection{Absence of global maximum of free energy}
\label{sec_global}

It is easy to show that there is no global maximum of free energy. To prove this result, we just need to consider a homogeneous sphere of mass $M$ and radius $R$. The total energy of this sphere is
\begin{equation}
E={3\over 2}NkT-{3GM^{2}\over 5R},
\label{Ehom}
\end{equation}
and its free energy
\begin{equation}
{J}={3\over 2}N k\ln\biggl ({2\pi kT\over m}\biggr )-Nk\ln\biggl ({3N\over 4\pi R^{3}}\biggr )+{3GM^{2}\over 5TR}.
\label{Jmax}
\end{equation}
From the above formula it is clear that $J$ goes to $+\infty$ when we spread the density to infinity ($R\rightarrow +\infty$) or when we contract the system to a point ($R\rightarrow 0$).  Since the mass is conserved in this process no global maximum of free energy can exist. 

In addition, there  is not even a local maximum of free energy in an unbounded domain: unbounded isothermal spheres have an {infinite mass}. For non rotating systems, the equilibrium states are expected to be spherically symmetric. In that case, the Boltzmann-Poisson equation (\ref{Mf})  can be rewritten
\begin{equation}
{1\over r^{2}}{d\over dr}\biggl (r^{2}{d\Phi\over dr}\biggr )=4\pi G A e^{-\beta \Phi}.  
\label{Mfr}
\end{equation} 
This equation has been studied extensively in the context of isothermal gaseous spheres (Chandrasekhar 1942). We can check by direct substitution that the distribution 
\begin{equation}
\Phi_{s} (r)={1\over \beta}\ln (2\pi G\beta A r^{2}),  \qquad \rho_{s}(r)= {1\over 2\pi G\beta r^{2}},
\label{singular}
\end{equation} 
is an exact solution of Eq. (\ref{Mfr}) known as the {\it singular isothermal sphere} (Binney \& Tremaine 1987). Since $\rho\sim r^{-2}$ at large distances, the total mass of the system $M=\int_{0}^{+\infty}\rho \ 4\pi r^{2}dr$ is infinite! More generally, we can show that any solution of the meanfield equation (\ref{Mfr}) behaves like the singular sphere as $r\rightarrow +\infty$ and has therefore an infinite mass  (Chandrasekhar 1942). 

We shall avoid the infinite mass problem by  confining artificially the system within a spherical box of radius $R$. It is only under this simplifying  assumption that a rigorous thermodynamics of self-gravitating systems can be carried out. Physically, the maximum cut-off determines the scale at which other processes intervene to limitate the spatial extent of the system. It is clear from Eq. (\ref{Jmax}) that the introduction of an upper cut-off does not remove the absence of global maximum of free energy (the system can always increase $J$ by collapsing to a point).  However, the presence of a wall at radius $R$ allows the existence of {\it local} maxima of free energy (metastable states).

\subsection{The equilibrium phase diagram}
\label{sec_diagram}

The microcanonical and canonical ensembles yield the same critical points, i.e. the critical points of entropy at fixed mass and energy and the critical points of free energy at fixed mass and temperature coincide. Only the onset of instability will differ from one ensemble to the other. The thermodynamical parameters for bounded isothermal spheres in the meanfield approximation have  been calculated by  Lynden-Bell \& Wood (1968) and we shall directly use their results.

\begin{figure}[htbp]
\centerline{
\psfig{figure=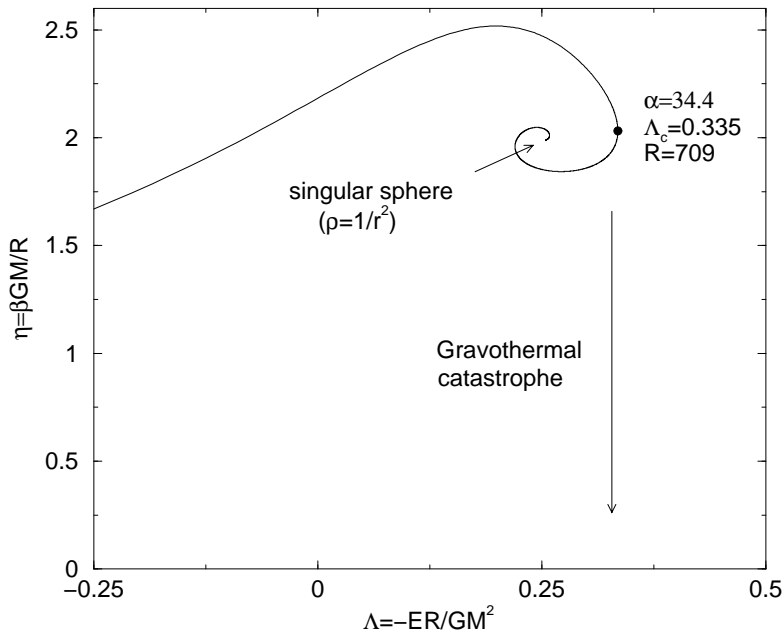,angle=0,height=8.5cm}}
\caption{Stability diagram for isothermal spheres in the microcanonical ensemble (fixed $E$).}
\label{gravotherme}
\end{figure}

\begin{figure}[htbp]
\centerline{
\psfig{figure=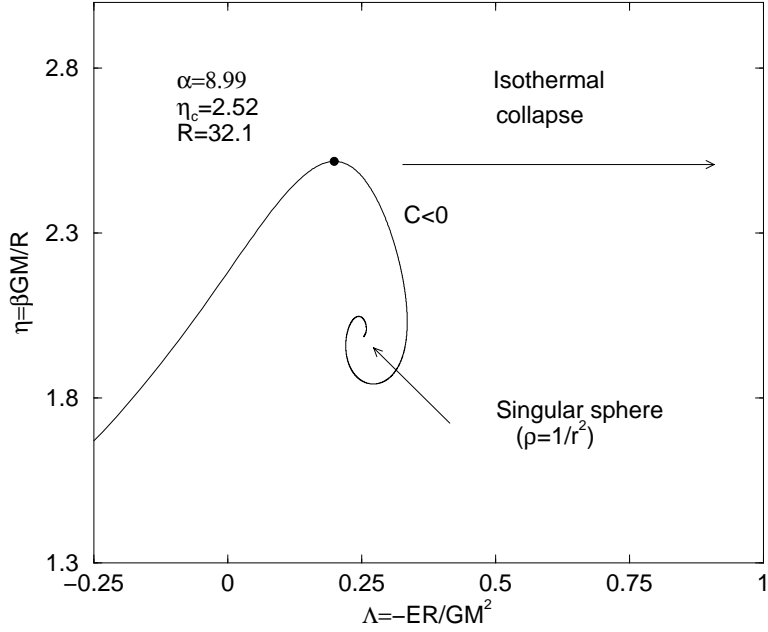,angle=0,height=8.5cm}}
\caption{Stability diagram for isothermal spheres in the canonical ensemble (fixed $T$).}
\label{isocollapse}
\end{figure}

To that purpose, we introduce the function  $\psi=\beta(\Phi-\Phi_{0})$ where $\Phi_{0}$ is the gravitational potential at $r=0$. Then, the density field can be written
\begin{equation}
\rho=\rho_{0}e^{-\psi},
\label{rhopsi}
\end{equation}
where $\rho_{0}$ is the central density. Introducing the notation $\xi=(4\pi G\beta\rho_{0})^{1/2}r$,  the Boltzmann-Poisson equation (\ref{Mfr}) reduces to the standard Emden form (Chandrasekhar 1942)
\begin{equation}
{1\over \xi^{2}}{d\over d\xi}\biggl (\xi^{2}{d\psi\over d\xi}\biggr )=e^{-\psi}.
\label{emden}
\end{equation}
 Eq. (\ref{emden}) has a simple analytic solution, the singular sphere
\begin{equation}
e^{-\psi_{s}}={2\over\xi^{2}}.
\label{sing2}
\end{equation}
The regular solutions of Eq.  (\ref{emden}) must satisfy the boundary conditions
\begin{equation}
\psi=\psi'=0,
\label{limxi0}
\end{equation}
when $\xi=0$. These solutions must be computed numerically. However, their asymptotic behaviors are well-known 
\begin{equation}
\psi={1\over 6}\xi^{2}-{1\over 120}\xi^{4}+{1\over 1890}\xi^{6}+...\qquad (\xi\rightarrow 0),
\label{ab1}
\end{equation}
\begin{equation}
e^{-\psi}={2\over\xi^{2}}\biggl \lbrace 1+ {A\over \xi^{1/2}}\cos\biggl ({\sqrt{7}\over 2}\ln\xi +\delta\biggr )\biggr \rbrace \quad (\xi\rightarrow +\infty).
\label{ab2}
\end{equation}
The curve (\ref{ab2}) intersects the singular solution (\ref{sing2}) infinitely often at points that asymptotically increase geometrically in the ratio $1:e^{2\pi/\sqrt{7}}=1:10.74...$. This property will have important physical consequences in section \ref{sec_frag}.

In the case of bounded isothermal systems, the solutions of Eq. (\ref{emden}) are terminated by the box at different radii given by  $\alpha=(4\pi G\beta\rho_{0})^{1/2}R$. Lynden-Bell \& Wood (1968) show that the reduced temperature and reduced energy can be expressed in terms of $\alpha$ by
\begin{equation}
\eta\equiv {\beta GM\over R}=\alpha\psi'(\alpha),
\label{eta}
\end{equation}
\begin{equation}
\Lambda\equiv -{ER\over GM^{2}}={3\over 2}{1\over \alpha\psi'(\alpha)}-{e^{-\psi(\alpha)}\over \psi'(\alpha)^{2}}.
\label{Lambda}
\end{equation}

For each value of the normalized inverse temperature ${\beta GM\over
R}$, we can solve Eq. (\ref{eta}) to get $\alpha$. Substituting the
result in Eq. (\ref{Lambda}), we deduce the corresponding value of the
normalized energy $-{ER\over GM^{2}}$. We can then determine the
equilibrium diagram $E-\beta$
(Figs. \ref{gravotherme}-\ref{isocollapse}). The critical points of
entropy or free energy form a spiral. This curve is parametrized by
$\alpha$ that goes from $0$ to $+\infty$ as we spiral inward. Instead
of the parameter $\alpha$, it may be more relevant to introduce the
density contrast
\begin{equation}
{\cal R}={\rho_{0}\over\rho(R)}=e^{\psi(\alpha)},
\label{contraste}
\end{equation}
which gives a more physical parametrization of the solutions. For
$\alpha\rightarrow 0$, we can use the asymptotic behavior $\psi\sim
{1\over 6}\alpha^{2}$ of the potential at the origin [Eq.
(\ref{ab1})] and we find ${\beta G M\over R}\sim {\alpha^{2}\over
3}\rightarrow 0$, $-{ER\over GM^{2}}\sim -{9\over
2\alpha^{2}}\rightarrow -\infty$ and ${\cal R}\rightarrow 1$. This
corresponds to high temperatures $T\rightarrow +\infty$. In this case,
self-gravity is negligible with respect to thermal motion and the
system behaves like an ordinary gas with uniform density. We check, by
eliminating $\alpha$ from the two foregoing relations, that
$E={3M\over 2\beta}$. This is indeed the equation of state for an
ideal gas without self-gravity. When we proceed along the spiral, the
density contrast increases monotonically and goes to $+\infty$ at the
center of the spiral. Using the asymptotic expansion
$e^{-\psi(\alpha)}\sim {2\over\alpha^{2}}$ of the potential at large
distances [Eq. (\ref{ab2})], we find ${\beta GM\over
R}\rightarrow 2$, $-{ER\over GM^{2}}\rightarrow {1\over 4}$ and ${\cal
R}\sim {\alpha^{2}\over 2}\rightarrow +\infty$. We can check by a
direct calculation that the ending point $(1/4,2)$ of the spiral
corresponds to the singular sphere (\ref{singular}).

\subsection{The Milne variables}
\label{sec_milne}

\begin{figure}[htbp]
\centerline{
\psfig{figure=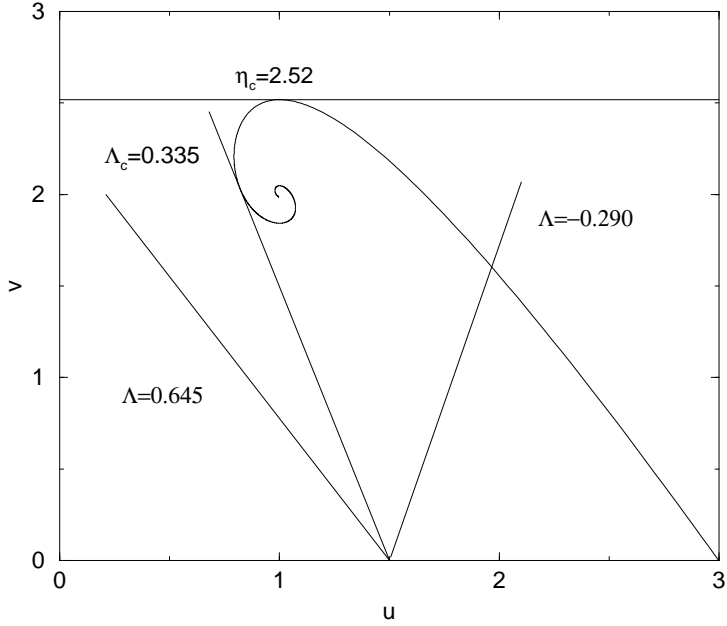,angle=0,height=8.5cm}}
\caption{The $(u,v)$ plane. All isothermal spheres must necessarily lie on the spiral. There exists solutions only for $\eta<2.52$ and $\Lambda<0.335$. }
\label{uvcrit}
\end{figure}

It will be convenient in the following to introduce the Milne variables $(u,v)$ defined by (Chandrasekhar 1942):
\begin{equation}
u={\xi e^{-\psi}\over \psi'} \qquad {\rm and }\qquad v=\xi \psi'.
\label{uv}
\end{equation}
Taking the logarithmic derivative of $u$ and $v$ with respect to $\xi$ and using Eq. (\ref{emden}), we get
\begin{equation}
{1\over u}{du\over d\xi}={1\over \xi}(3-v-u), 
\label{uv1}
\end{equation}
\begin{equation}
{1\over v}{dv\over d\xi}={1\over \xi}(u-1). 
\label{uv2}
\end{equation}
Taking the ratio of the foregoing equations, we find that the variables $u$ and $v$ are related to each other by a first order differential equation 
\begin{equation}
{u\over v}{dv\over du}=-{u-1\over u+v-3}.
\label{uv3}
\end{equation}
Therefore, by using the Milne variables, the degree of the meanfield equation (\ref{emden}) has been reduced from two to one. As discussed extensively  by Chandrasekhar (1942), this property is related to the homology invariance of the solutions of the Emden equation. The solution curve in the $(u,v)$ plane is well known and is ploted in Fig. \ref{uvcrit}.  The curve is parametrized by $\xi$. Starting from the point $(u,v)=(3,0)$ corresponding to $\xi=0$, the solution curve spirals indefinitely around the point $(u_{s},v_{s})=(1,2)$, corresponding to the singular sphere, as $\xi$ tends to infinity. All isothermal spheres must necessarily lie on this curve. For bounded isothermal spheres, $\xi$ must be terminated at the box radius $\alpha$. Clearly, the spiral behavior of the $(u,v)$ curve (and also the curve of Figs. \ref{gravotherme}-\ref{isocollapse}) can be ascribed to the oscillating behavior of the solution (\ref{ab2}) as $\xi\rightarrow +\infty$.

It turns out that the  normalized temperature and energy can be expressed very simply in terms of the values of $u$ and  $v$ at the normalized box radius $\alpha$. Indeed, writing $u_{0}=u(\alpha)$ and  $v_{0}=v(\alpha)$ and using Eqs. (\ref{eta})-(\ref{Lambda}), we get
\begin{equation}
\Lambda={1\over v_{0}}\biggl ({3\over 2}-u_{0}\biggr ),
\label{uvLambda}
\end{equation} 
\begin{equation}
\eta=v_{0}.
\label{uvbeta}
\end{equation}
The relation (\ref{uvLambda}) was previously noticed by Padmanabhan (1989). The intersection between the lines defined by Eqs. (\ref{uvLambda})-(\ref{uvbeta}) and the spiral in the $(u,v)$ plane determines the value of $\alpha$ corresponding to a given energy or temperature. As noticed by Padmanabhan, for  $\Lambda>\Lambda_{c}=0.335$ there is no intersection. Thus, in the microcanonical ensemble, no isothermal sphere can exist if $-{ER\over GM^{2}}>0.335$. This result can also be read from Fig. \ref{gravotherme} and was discovered by Antonov (1962). For sufficiently small energies, the system is expected to collapse and overheat in agreement with the ``gravothermal catastrophe'' picture (Lynden-Bell \& Wood 1968). Similarly, considering Eq. (\ref{uvbeta}), we find that there is no intersection for ${\beta GM\over R}>v_{max}=2.52$ (where $v_{max}$ corresponds to $u=1$). This result can also be read from Fig. \ref{isocollapse}. In the canonical ensemble, a gaseous sphere is expected to collapse below a critical temperature $kT_{c}={GmM\over 2.52 R}$ (Lynden-Bell \& Wood 1968).

\subsection{The stability analysis of Katz}
\label{sec_katz}

More precisely, it is possible to prove the following stability result: when a self-gravitating system is confined within a box and maintained at a constant  temperature $T$, local maxima of free energy exist only for  ${\beta GM\over R}\le 2.52$; they have a density contrast ${\cal R}=\rho(0)/\rho(R)<32.2$ (and $\alpha<8.99$). Critical points of free energy with density contrast ${\cal R}>32.2$ (and $\alpha>8.99$) are unstable saddle points. For ${\beta GM\over R}> 2.52 $, there are no critical points of free energy. 

The stability of the solutions can be deduced from the topology of the $E-\beta$ curve by using the method of Katz (1978) who has extended Poincar\'e's theory of linear series of equilibrium. The parameter conjugate to the free energy with respect to the inverse temperature $\beta$ is $-E=({\partial J\over\partial \beta})_{M,R}$. Then, if we plot $E$ as a function of $\beta$ (we just need to rotate Fig. \ref{isocollapse} by $90^{o}$) we have the following results: (i) a change of stability can occur only at a limit point where $\beta$ is an extremum ($dE/d\beta$ infinite) (ii) a mode of stability is lost when the curve rotates clockwise and gained otherwise. Now, we know that for $T$ sufficiently large the solutions are stable because, in this limit, self-gravity is negligible and the system behaves like an ordinary gas. From point (i) we conclude that the solutions with  ${\cal R}=\rho(0)/\rho(R)<32.2$  are stable. As the curve spirals inward for ${\cal R}>32.2$, more and more modes of stability are lost. In this respect, the singular sphere, at the end of the spiral, is the most unstable solution \footnote{Another application of Katz theorem where the spiral unwinds and the stability is regained is given by Chavanis \& Sommeria (1998) for self-gravitating fermions.}. It can be noted that instability sets in precisely when the specific heat $C=dE/dT$ becomes negative (in the range $32.2<{\cal R}<709$). General considerations indicate that negative specific heats are forbidden in the canonical ensemble (Padmanabhan 1990). By contrast, thermally isolated systems can have negative specific heats: in the microcanonical ensemble, the solutions with $32.2<{\cal R}<709$ are stable (Katz 1978, Padmanabhan 1989). This clearly demonstrates  that the statistical ensembles do not coincide for self-gravitating systems: the region of negative specific heats in the microcanonical ensemble is replaced by a phase transition (an ``isothermal collapse'') in the canonical ensemble (Padmanabhan 1990). It is at the verge of this phase transition that the meanfield approximation ceases to be valid as demonstrated by de Vega {\it et al.} (2001a, 2001b).

\subsection{Analogy with critical phenomena}
\label{sec_critical}

In this section, we study some analogies with the theory of critical
phenomena. Further analogies are discussed in Chavanis {\it et al.}
(2001) where a simple dynamical model for studying phase transitions
in self-gravitating systems is proposed.

\begin{figure}[htbp]
\centerline{
\psfig{figure=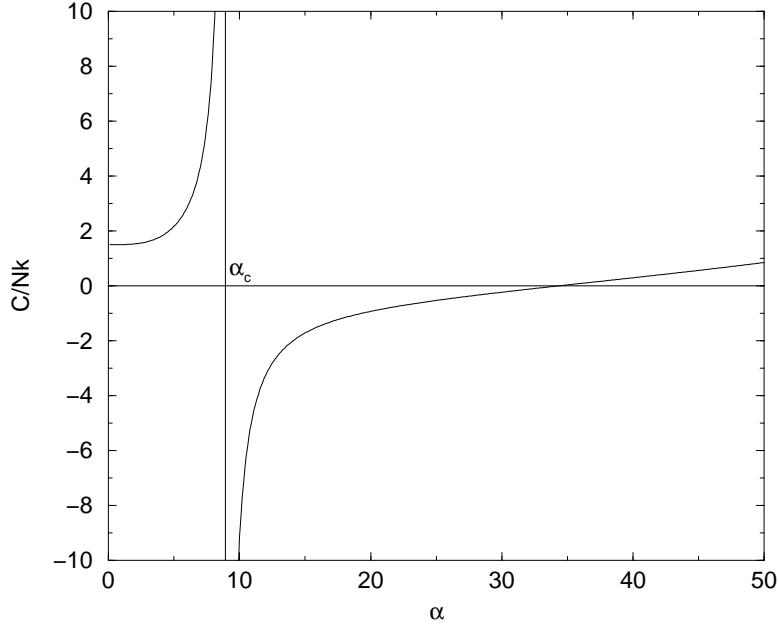,angle=0,height=8.5cm}}
\caption{Evolution of the specific heat as a function of $\alpha$. The specific heat diverges for the first time at $\alpha_{c}=8.99$ and is negative in the range $8.99<\alpha<34.4$. For $\alpha=0$, $C={3\over 2}Nk$ (perfect gas).}
\label{capacite}
\end{figure}

The specific heat can be written 
\begin{equation}
C={dE\over dT}=-{k\over m}\beta^{2}{dE\over d\beta}=Nk\eta^{2}{d\Lambda\over d\eta}.
\label{crit1}
\end{equation} 
Using Eqs. (\ref{uvLambda})(\ref{uvbeta})(\ref{uv1})(\ref{uv2}), we easily get 
\begin{equation}
{d\eta\over d\alpha}={v_{0}\over\alpha}(u_{0}-1),
\label{crit2}
\end{equation} 
\begin{equation}
{d\Lambda\over d\alpha}={1\over 2\alpha v_{0}}(4u_{0}^{2}+2u_{0}v_{0}-11u_{0}+3)
\label{crit3}
\end{equation} 
Therefore, the specific heat is determined as a function of $\alpha$ by the expression
\begin{equation}
C=Nk {4u_{0}^{2}+2u_{0}v_{0}-11u_{0}+3\over 2(u_{0}-1)}.
\label{crit4}
\end{equation} 
Of course, $C\rightarrow \infty$ if $d\eta=0$ and $C\rightarrow 0$ if $d\Lambda=0$. On the other hand, for the ideal gas $(u_{0},v_{0})=(3,0)$, we recover the well-known result $C={3\over 2}Nk$. The specific heat is ploted as a function of $\alpha$ on Fig. \ref{capacite}.

\begin{figure}[htbp]
\centerline{
\psfig{figure=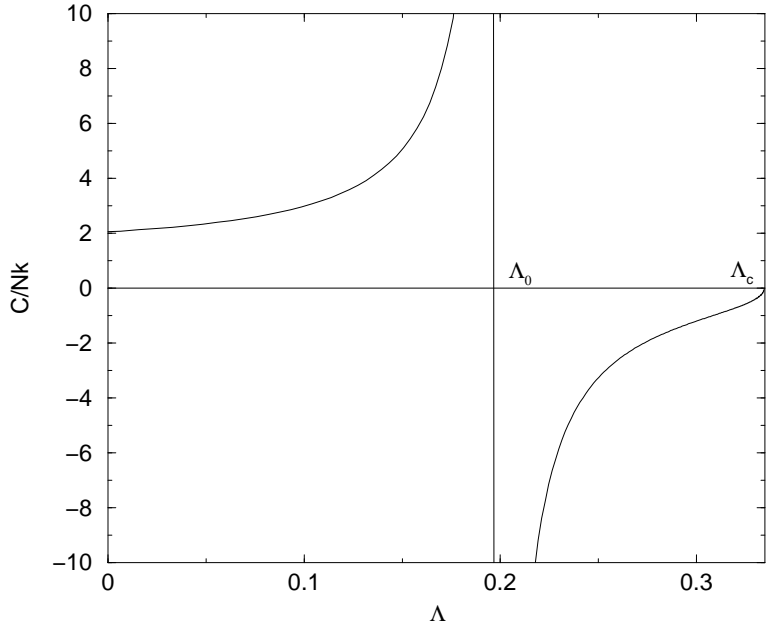,angle=0,height=8.5cm}}
\caption{Evolution of the specific heat as a function of $\Lambda$ near the point $\Lambda_{0}$. The specific heat diverges like $C\sim -(\Lambda-\Lambda_{0})^{-1}$. It is negative for $\Lambda>\Lambda_{0}$ and tends to ${3\over 2}Nk$ for $\Lambda\rightarrow -\infty$ (high energies). These results are supported by the numerical simulations of Cerruti-Sola {\it et al.} (2001) in the microcanonical ensemble who solved the gravitational $N$-body problem in a finite container.}
\label{phencrit}
\end{figure}

\begin{figure}[htbp]
\centerline{
\psfig{figure=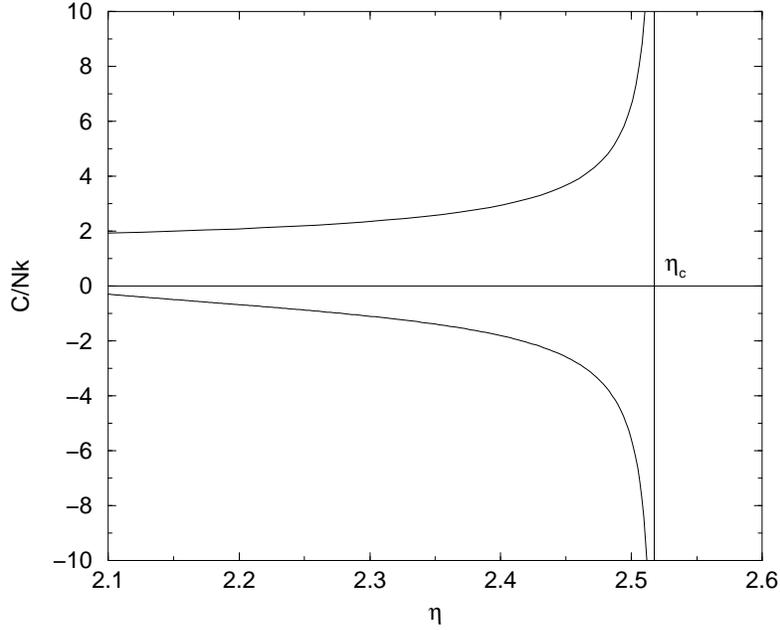,angle=0,height=8.5cm}}
\caption{Evolution of the specific heat as a function of $\eta$ near the critical point $\eta_{c}$. The specific heat diverges like $C\sim \pm (\eta_{c}-\eta)^{-1/2}$. It tends to ${3\over 2}Nk$ for $\eta\rightarrow 0$ (high temperatures). The region of negative specific heat (lower curve) is forbidden in the canonical ensemble. }
\label{Ceta}
\end{figure}

At the critical point $\eta=\eta_{c}$ at which $C$ diverges, we write $\alpha=\alpha_{c}+\epsilon$ and expand $\eta$ and $\Lambda$ in terms of $\epsilon$, using Eqs. (\ref{crit2})-(\ref{crit3}) and $u_{0}(\alpha_{c})=1$, $v_{0}(\alpha_{c})=\eta_{c}$. To lowest order, we have
\begin{equation}
\eta-\eta_{c}=\biggl ({d^{2}\eta\over d\alpha^{2}}\biggr )_{\alpha_{c}}{\epsilon^{2}\over 2}={\eta_{c}\over 2\alpha_{c}^{2}}(2-\eta_{c})\epsilon^{2},
\label{crit5}
\end{equation} 
\begin{equation}
\Lambda-\Lambda_{0}=\biggl ({d\Lambda\over d\alpha}\biggr )_{\alpha_{c}}\epsilon={\eta_{c}-2\over \alpha_{c}\eta_{c}}\epsilon,
\label{crit6}
\end{equation} 
where $\Lambda_{0}\simeq 0.197$ is the value of $\Lambda$ at $\alpha=\alpha_{c}$. Eliminating $\epsilon$ from these expressions, we find the relation between $\eta$ and $\Lambda$ close to the critical point
\begin{equation}
\eta_{c}-\eta={\eta_{c}^{3}\over 2(\eta_{c}-2)}(\Lambda-\Lambda_{0})^{2}.
\label{crit7}
\end{equation} 
Then, using Eq. (\ref{crit1}), we find explicitely
\begin{eqnarray}
C=\Biggl\lbrace \begin{array}{cc}
-Nk{\eta_{c}-2\over \eta_{c}}(\Lambda-\Lambda_{0})^{-1},
 \\ \pm Nk\biggl \lbrack {\eta_{c}(\eta_{c}-2)\over 2}\biggr \rbrack^{1/2}(\eta_{c}-\eta)^{-1/2}.
\end{array}
\label{crit8}
\end{eqnarray} 
The specific heat diverges as a function of the energy with an exponent $\nu=1$ and as a function of the temperature with an exponent $\nu'=1/2$ (see Figs. \ref{phencrit}-\ref{Ceta}). It becomes {\it negative} for $\Lambda_{0}<\Lambda<\Lambda_{c}$. This region of negative specific heats is allowed in the microcanonical ensemble but not in the canonical ensemble.

It is also of interest to determine the behavior of the central density $\rho_{0}$ near the critical point (C. Sire, private communication). The central density is related to $\alpha$ by the relation $\alpha=(4\pi G\beta\rho_{0})^{1/2}R$ or, alternatively,
\begin{equation}
\rho_{0}={\alpha^{2}M\over 4\pi R^{3}\eta}.
\label{crit9}
\end{equation} 
To first order in $\epsilon$, we have
\begin{equation}
\rho_{0}-\rho_{0}^{c}={2\alpha_{c}M\over 4\pi R^{3}\eta_{c}}\epsilon.
\label{crit10}
\end{equation} 
where $\rho_{0}^{c}=\alpha_{c}^{2}M/(4\pi R^{3}\eta_{c})$ is the value of the central density at the critical point $\eta_{c}$.  Using Eqs. (\ref{crit5})-(\ref{crit6}), we get
\begin{eqnarray}
{\rho_{0}^{c}-\rho_{0}\over \rho_{0}^{c}}=\Biggl\lbrace \begin{array}{cc}
-{2\eta_{c}\over\eta_{c}-2}(\Lambda-\Lambda_{0}),
 \\ \pm \biggl\lbrack {8\over \eta_{c}-2}\biggl (1-{\eta\over\eta_{c}}\biggr )\biggr \rbrack^{1/2}.
\end{array}
\label{crit11}
\end{eqnarray}
Quite remarkably, this stationary result can also be derived from dynamical considerations (see Chavanis {\it et al.} 2001). Note also that, using Eqs. (\ref{contraste})(\ref{uv}), we find ${\cal R}_{c}=\alpha_{c}^{2}/\eta_{c}$, which implies $\rho^{c}(R)=M/(4\pi R^{3})$.

\section{An extension of Padmanabhan's stability analysis}
\label{sec_pad}

\subsection{The condition of thermodymamical stability  }
\label{sec_cond}

The stability of the gaseous spheres can be studied by extending Padmanabhan (1989)'s stability analysis to the canonical ensemble. This will provide the profiles of the perturbations that induce instability. 

A critical point of free energy is a local maximum if, and only if, the second variations 
\begin{equation}
\delta^{2}J=-{k\over m}\int {(\delta\rho)^{2}\over 2\rho} d^{3}{\bf r}-{1\over 2T}\int \delta\rho\delta\Phi d^{3}{\bf r},
\label{s1}
\end{equation}
are negative for any variation $\delta\rho$ that conserves mass to first order.
This conservation law imposes
\begin{equation}
\int \delta\rho d^{3}{\bf r}=0.
\label{s2}
\end{equation}
We shall restrict ourselves to spherically symmetric perturbations since only  spherically symmetric perturbations can induce instability for non rotating systems. Following Padmanabhan (1989), we introduce the variable $q$ defined by
\begin{equation}
\delta\rho={1\over 4\pi r^{2}}{dq\over dr}.
\label{s3}
\end{equation}
Physically, $q$ represents the mass perturbation $q(r)\equiv \delta M(r)=\int_{0}^{r}4\pi r^{'2}\delta\rho(r')dr'$ within the sphere of radius $r$. The boundary conditions on $q$ are thus
\begin{equation}
q(0)=q(R)=0.
\label{s4}
\end{equation}
Substituting the foregoing expression for  $q(r)$ in Eq. (\ref{s1}), we obtain 
\begin{equation}
\delta^{2}J=-{1\over 2T}\int_{0}^{R} {dq\over dr}\delta \Phi dr-{k\over m}\int_{0}^{R} {1\over 8\pi\rho r^{2}}\biggl ({dq\over dr}\biggr )^{2} dr.
\label{s5}
\end{equation}
Integrating by parts and using the boundary conditions (\ref{s4}), we get
\begin{equation}
\delta^{2}J={1\over 2T}\int_{0}^{R} q {d\delta \Phi\over dr} dr+{k\over m}\int_{0}^{R} q {d\over dr}\biggl ({1\over 8\pi\rho r^{2}}{dq\over dr}\biggr ) dr.
\label{s6}
\end{equation}
Then, using Gauss theorem in the form 
\begin{equation}
{d\delta\Phi\over dr}={Gq\over r^{2}},
\label{s7}
\end{equation}
we find
\begin{equation}
\delta^{2}J={1\over 2T}\int_{0}^{R} {Gq^{2}\over r^{2}} dr+{k\over m}\int_{0}^{R} q {d\over dr}\biggl ({1\over 8\pi\rho r^{2}}{dq\over dr}\biggr ) dr,
\label{s8}
\end{equation}
or, equivalently,
\begin{equation}
\delta^{2}J={1\over 2}\int_{0}^{R}dr q\biggl\lbrack  {G\over Tr^{2}} +{k\over m} {d\over dr}\biggl ({1\over 4\pi\rho r^{2}}{d\over dr}\biggr )\biggr\rbrack q.
\label{s9}
\end{equation}
The second variations of free energy can be positive (implying instability) only if the differential operator which occurs in the integral has positive eigenvalues.  We need therefore to consider the eigenvalue problem 
\begin{equation}
\biggl\lbrack {k\over m} {d\over dr}\biggl ({1\over 4\pi\rho r^{2}}{d\over dr}\biggr )+ {G\over Tr^{2}}\biggr\rbrack q_{\lambda}(r)=\lambda q_{\lambda}(r),
\label{s10}
\end{equation}
with $ q_{\lambda}(0)=q_{\lambda}(R)=0$. If all the eigenvalues $\lambda$ are negative, then the critical point is a {\it maximum } of free energy. If at least one eigenvalue is positive, the critical point is an unstable saddle point. The point of marginal stability $\eta=\eta_{c}$ is determined by the condition that the largest eigenvalue is equal to zero. We thus have to solve the differential equation
\begin{equation}
 {k\over m} {d\over dr}\biggl ({1\over 4\pi\rho r^{2}}{dF\over dr}\biggr )+ {GF(r)\over Tr^{2}}=0,
\label{s11}
\end{equation}
with $F(0)=F(R)=0$. Note that this equation corresponds to that found by Padmanabhan (1989) in the microcanonical ensemble with $V=0$.

\subsection{The point of marginal stability}
\label{sec_marginal}

\begin{figure}[htbp]
\centerline{
\psfig{figure=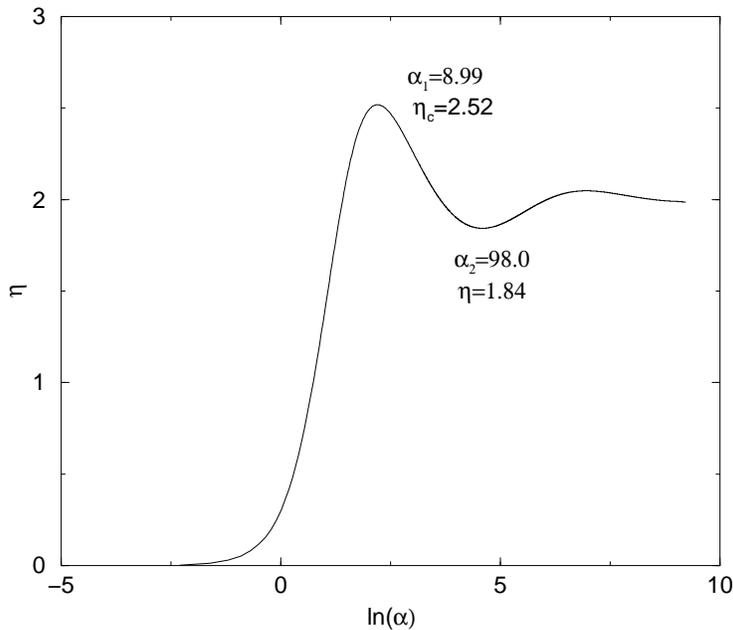,angle=0,height=8.5cm}}
\caption{Evolution of the inverse temperature $\eta$ as a function of the scaled radius $\alpha$. A mode of stability is lost each time that $\eta$ is extremum. This figure is the counterpart of Fig. 3 of Padmanabhan (1989) in the microcanonical ensemble.  }
\label{etaalpha}
\end{figure}

 Introducing the dimensionless variables of section \ref{sec_diagram}, we can rewrite the foregoing equation in the form
\begin{equation}
{d\over d\xi}\biggl ({e^{\psi}\over \xi^{2}}{dF\over d\xi}\biggr )+ {F(\xi)\over \xi^{2}}=0,
\label{s12}
\end{equation}
with $F(0)=F(\alpha)=0$. As shown by Padmanabhan (1989), this equation can be solved without solving explicitly the Emden equation (\ref{emden}). Let us introduce the differential operator
\begin{equation}
{\cal L}= {d\over d\xi}\biggl ({e^{\psi}\over \xi^{2}}{d\over d\xi}\biggr )+ {1\over \xi^{2}}.
\label{s13}
\end{equation}
Using Eq. (\ref{emden}), it is readily established that 
\begin{eqnarray}
{\cal L}(\xi^{2}\psi')={d\over d\xi}\biggl ({e^{\psi}\over \xi^{2}}{d\over d\xi}(\xi^{2}\psi')\biggr )+ \psi'=
{d\over d\xi}(e^{\psi}\times e^{-\psi})+\psi'=\psi'.
\label{s13bis}
\end{eqnarray}
Similarly, we find
\begin{eqnarray}
{\cal L}(\xi^{3}e^{-\psi})={d\over d\xi}\biggl ({e^{\psi}\over \xi^{2}}{d\over d\xi}(\xi^{3}e^{-\psi})\biggr )+ \xi e^{-\psi}\nonumber\\
={d\over d\xi}(3-\xi\psi')+ \xi e^{-\psi}=-\xi\psi''-\psi'+ \xi e^{-\psi}=\psi'.
\label{s14}
\end{eqnarray}
These identities suggest to seek  a solution of Eq. (\ref{s12}) in the form
\begin{equation}
F(\xi)=c_{1}\xi^{3}e^{-\psi}+c_{2}\xi^{2}\psi'
\label{s16}
\end{equation}
where $c_{1}$ and  $c_{2}$ are constants. Substituting the foregoing expression for $F(\xi)$ in the differential equation  (\ref{s12}), we find that $c_{1}+c_{2}=0$. The function  $F(\xi)$ can therefore be reexpressed as 
\begin{equation}
F(\xi)=c_{1} (\xi^{3}e^{-\psi}-\xi^{2}\psi').
\label{s17}
\end{equation}
The boundary condition $F(0)=0$ is automatically satisfied (we can show furthermore that Eq. (\ref{s17}) is the only solution consistent with this boundary condition). The condition $F(\alpha)=0$ will determine the point on the spiral of Fig. \ref{isocollapse} at which the gaseous spheres become unstable (i.e., the point of marginal stability). From Eq. (\ref{s17}), we get
\begin{equation}
\alpha^{3}e^{-\psi(\alpha)}-\alpha^{2}\psi'(\alpha)=0.
\label{s18}
\end{equation}
This can be expressed in terms of the Milne variables (\ref{uv}) as  
\begin{equation}
u_{0}=1.
\label{s19}
\end{equation}
Considering Eq. (\ref{uv3}), we see that the condition (\ref{s19}) determines the points for which $v$ is extremum ($dv/du=0$). Since $\eta=v_{0}$ according to Eq. (\ref{uvbeta}), we deduce that a change of stability occurs each time that the temperature is extremum (see Fig. \ref{etaalpha}) in complete agreement with the theorem of Katz (1978). In particular, the first instability (for which the largest eigenvalue $\lambda$ becomes positive) corresponds to $\eta_{c}=v_{max}=2.52$ (for this value $\alpha=8.99$). We show in the next section that the secondary instabilities occur at values of $\alpha$ that follow asymptotically a geometric progression. The profiles of the perturbations that induce these instabilities are discussed in section \ref{sec_profile}.

\subsection{Secondary instabilities}
\label{sec_secinst}

A mode of stability is lost each time that $u_{0}=u(\alpha)=1$. Now, it is easy to have an analytical estimate of $u(\xi)$ for large values of $\xi$. Introducing the asymptotic behavior (\ref{ab2}) of $\psi$ in the Milne variable $u$ defined by Eq. (\ref{uv}) we get for $\xi\rightarrow +\infty$:
\begin{eqnarray}
u={1+ {A\over \xi^{1/2}}\cos\bigl ({\sqrt{7}\over 2}\ln\xi +\delta\bigr )\over 1+ {A\over 4\xi^{1/2}}\bigl\lbrack \sqrt{7}\sin\bigl ({\sqrt{7}\over 2}\ln\xi +\delta\bigr )+\cos\bigl ({\sqrt{7}\over 2}\ln\xi +\delta\bigr )\bigr\rbrack}.
\label{g2}
\end{eqnarray}
Therefore, the condition $u_{0}=1$ corresponds to values 
of $\alpha$ satisfying 
\begin{equation}
\tan\biggl ({\sqrt{7}\over 2}\ln\alpha +\delta\biggr )={3\over\sqrt{7}}\qquad (\alpha\rightarrow +\infty),
\label{g3}
\end{equation}
or equivalently
\begin{eqnarray}
{\sqrt{7}\over 2}\ln\alpha_{n} +\delta =\arctan (3/\sqrt{7})+n\pi  \quad (n \ {\rm integer}).
\label{g3bis}
\end{eqnarray}
This determines a succession of values that follow the geometric progression
\begin{equation}
\alpha_{n} \sim e^{2\pi n\over\sqrt{7}}=[10.74...]^{n} \qquad (n\rightarrow +\infty,\ {\rm integer}).
\label{g4}
\end{equation}
At these points a new eigenvalue becomes positive implying secondary instabilities. Clearly, this geometric progression can be traced back to the `curious' asymptotic behavior of the solutions governing the isothermal gas sphere (see Eq. (\ref{ab2})) which intersects the singular solution (\ref{sing2}) infinitely often at points that asymptotically increase geometrically in the ratio $1:10.74$. 

A mode of stability is lost each time that $\alpha$ achieves one of the values given by Eq. (\ref{g4}). If one prefers, the same criterion can be expressed in terms of the density contrast ${\cal R}$. Using Eq. (\ref{contraste}) and the asymptotic expansion (\ref{ab2}) we get ${\cal R}=e^{\psi(\alpha)}\sim {\alpha^{2}\over 2}$ for $\alpha\rightarrow +\infty$. Hence, the result (\ref{g4}) is translated in
\begin{equation}
{\cal R}_{n}\sim  e^{4\pi n\over\sqrt{7}}=[115.5...]^{n} \qquad (n\rightarrow +\infty,\ {\rm integer}). 
\label{g6}
\end{equation}

\subsection{Profiles of the perturbation at the critical points}
\label{sec_profile}

In this section, we study the form of the perturbations that trigger instability at the critical points. Following the method of Padmanabhan (1989), it is possible to describe the behavior of these perturbation profiles without numerical work. Indeed, according to Eq. (\ref{s3}), the eigenfunction associated with the eigenvalue $\lambda=0$ can be written
\begin{equation}
{\delta\rho\over\rho_{0}}={1\over 4\pi \xi^{2}}{dF\over d\xi}
\label{P1}
\end{equation}
where $F(\xi)$ is given by Eq. (\ref{s17}). Substituting this result in Eq. (\ref{P1}) and simplifying the derivative with the aid of Eq. (\ref{emden}) we obtain
\begin{equation}
{\delta\rho\over\rho}={c_{1}\over 4\pi}(2-\xi\psi').
\label{P2}
\end{equation}
Introducing the Milne variable $v$ defined by Eq. (\ref{uv}), we can rewrite the foregoing equation in the form
\begin{equation}
 {\delta\rho\over\rho}={c_{1}\over 4\pi} (2-v).
\label{P3}
\end{equation}

\begin{figure}[htbp]
\centerline{
\psfig{figure=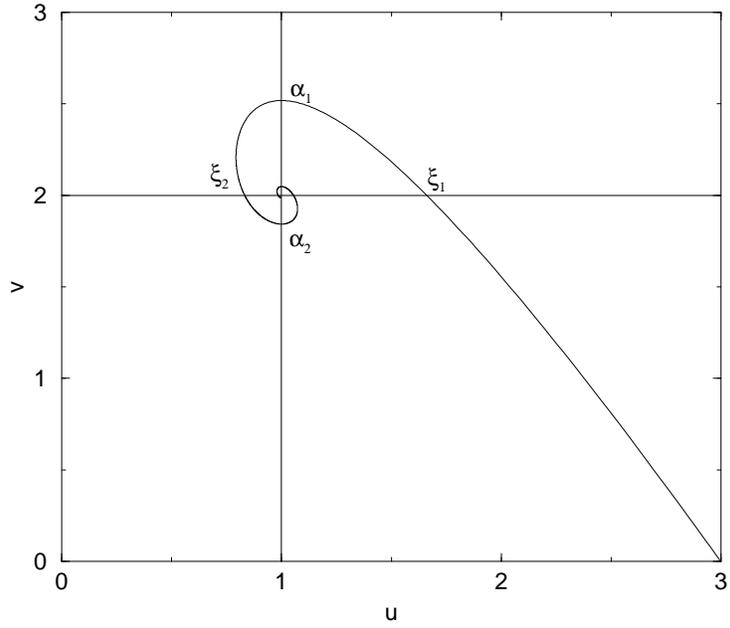,angle=0,height=8.5cm}}
\caption{Position of the zeros of the perturbation profile $\delta\rho/\rho$ in the $(u,v)$ plane.}
\label{uvosci}
\end{figure}

\begin{figure}[htbp]
\centerline{
\psfig{figure=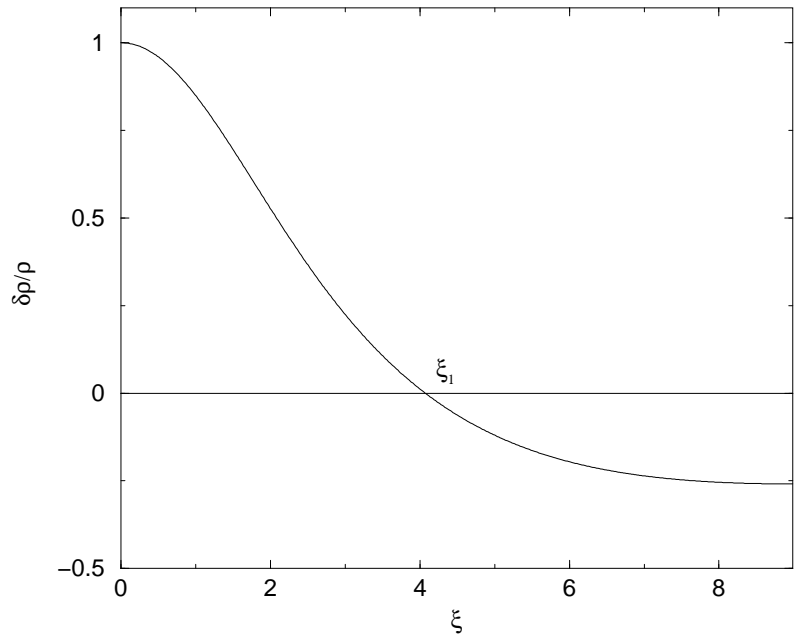,angle=0,height=8.5cm}}
\caption{First mode of instability corresponding to $\alpha_{1}=8.99$.}
\label{dnsn-xb2.51760387XI}
\end{figure}

\begin{figure}[htbp]
\centerline{
\psfig{figure=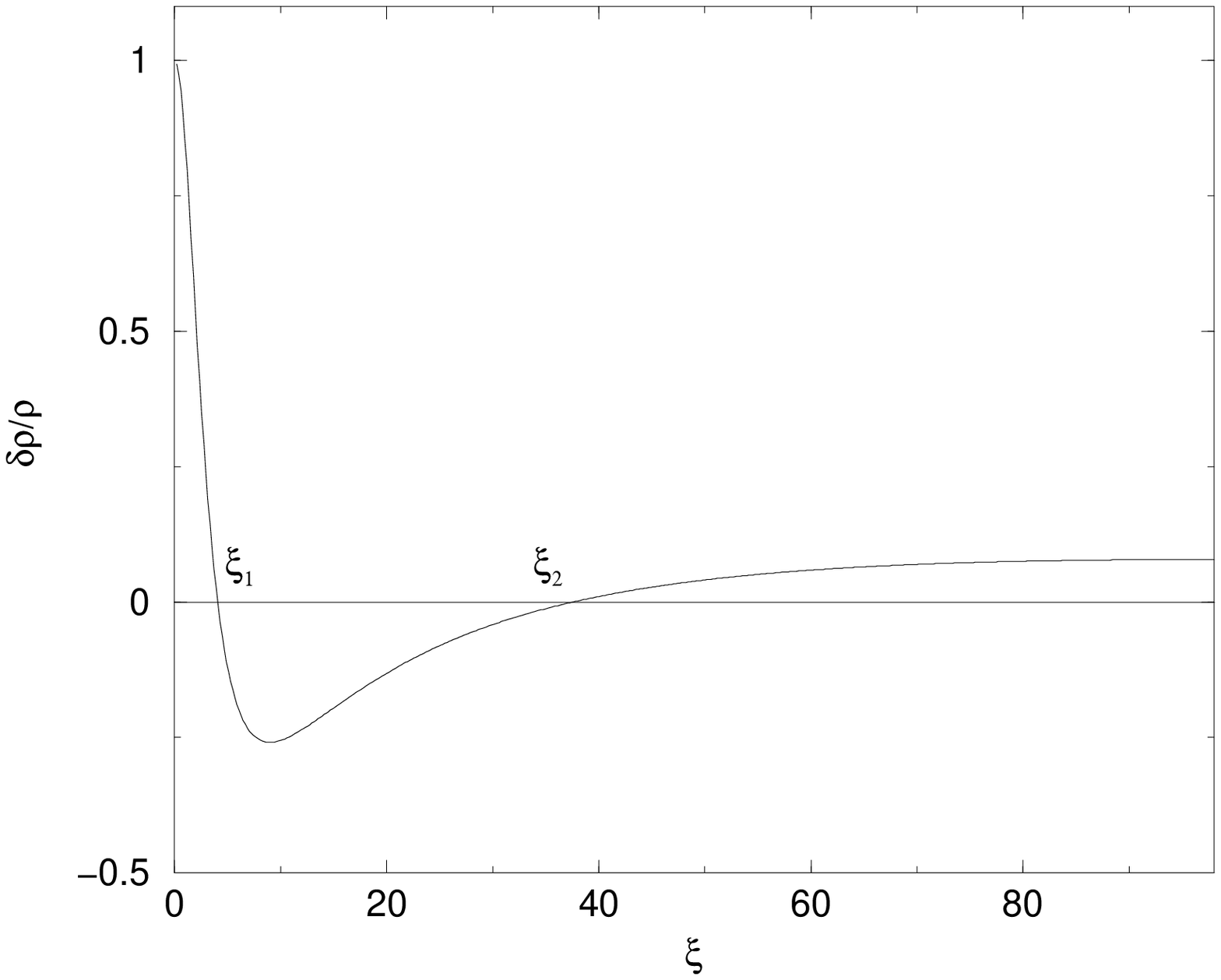,angle=0,height=8.5cm}}
\caption{Second mode of instability corresponding to $\alpha_{2}=98.0$. }
\label{dnsn-x.b1.84269667XI}
\end{figure}

The density perturbation $\delta\rho$ becomes zero at the point(s)
$\xi_{i}$ such that $v(\xi_{i})=2$.  The number of zeros is therefore
given by the number of intersections between the spiral in the $(u,v)$
plane and the line $v=2$ (see Fig. \ref{uvosci}). For the first mode
of instability $(\alpha_{1}=8.99)$, we need to follow the spiral until
the first extremum of $v$, which corresponds to the point at which the
box terminates (recall that $u(\alpha_{1})=1$, see section
\ref{sec_marginal}). In that case there is only one intersection
$\xi_{1}$ with the line $v=2$. Therefore, the first mode of
instability does {\it not} present a ``core-halo'' structure
(Fig. \ref{dnsn-xb2.51760387XI}), contrary to what happens in the
microcanonical ensemble (Padmanabhan 1989). For the second mode of
instability ($\alpha_{2}=98.0$), we need to follow the spiral until
the second extremum of $v$. The intersection with the line $v=2$ now
determines two zeros $\xi_{1}$ and $\xi_{2}$, the first one being the
same as for the first mode of instability. The profile of the second
mode of instability therefore has a ``core-halo'' structure
(Fig. \ref{dnsn-x.b1.84269667XI}). When we spiral inward, more and
more intersections are obtained. Therefore the high order modes of
instability present more and more oscillations. For these modes, it is
easy to determine the asymptotic positions of their zeros. From
Eqs. (\ref{uv}) and (\ref{ab2}), we have for $\xi\rightarrow +\infty$:
\begin{eqnarray}
v=2+{A\over 2\xi^{1/2}}
\times\biggl \lbrack \sqrt{7}\sin\biggl ({\sqrt{7}\over 2}\ln\xi+\delta\biggr )
+\cos\biggl ({\sqrt{7}\over 2}\ln\xi+\delta\biggr )\biggr \rbrack.
\label{P4}
\end{eqnarray}
Thus, substituting $v=2$ in Eq. (\ref{P4}), we get  
\begin{eqnarray}
{\sqrt{7}\over 2}\ln\xi_{i}+\delta=-\arctan(1/\sqrt{7})+i\pi \quad (i \ {\rm integer}).
\label{P4bis}
\end{eqnarray}
Therefore, the zeros follow asymptotically the geometric progression \footnote{It might be recalled, parenthetically, that the positions of the planets in the solar system also follow a geometric progression, but with a different ratio $q\simeq 2$ (Titius-Bode law). This is recognized to be an effect of scale invariance (Graner \& Dubrulle 1994, Dubrulle \& Graner 1994). There is, however, no real connexion with the present work and planets may have formed quite differently (see Chavanis (2000) and references therein).}
\begin{equation}
\xi_{i}\sim e^{2\pi i\over\sqrt{7}}=[10.74...]^{i} \qquad (i\rightarrow +\infty, i \ {\rm integer}).
\label{P5}
\end{equation}
Note that the perturbation that destabilizes the singular sphere (at the end of the spiral) has an {\it infinite} number of oscillations!

\section{Dynamical stability of isothermal gaseous spheres}
\label{sec_dyn}

We shall now investigate the dynamical stability of isothermal spheres described by Navier-Stokes equations and compare the results with the thermodynamical approach. This problem was previously addressed by Yabushita (1968) in the isobaric ensemble and by  Semelin {\it et al.} (2001) in the canonical ensemble. We shall prove  analytically and with no approximation the equivalence between dynamical and thermodynamical stability and the fact that the stability of isothermal spheres does not depend on the viscosity. We shall also describe the behavior of the velocity profiles at the critical points.

The equations of the problem are the equation of continuity, the equation of motion and the Poisson equation  
\begin{equation}
{\partial \rho\over \partial t}+\nabla (\rho {\bf v})=0,
\label{F1}
\end{equation}
\begin{eqnarray}
{\partial {\bf v}\over \partial t}+({\bf v}\nabla){\bf v}=-{1\over\rho}\nabla p-\nabla\Phi+{\eta\over\rho}\Delta {\bf v}+{1\over\rho}(\zeta+{\eta\over 3})\nabla(\nabla {\bf v}),
\label{F2}
\end{eqnarray}
\begin{equation}
\Delta\Phi=4\pi G\rho.
\label{F3}
\end{equation}
They must be completed by an equation of state that we take of the form $p=\rho {k\over m}T$ where $T$ is constant. We also recall that we work within a box of radius $R$. For astrophysical fluids, the Reynolds numbers are so huge that we can neglect the viscosity in the Navier-Stokes equation. We shall take therefore $\eta=\zeta=0$ in most of the calculations but we will also prove that viscosity does not change the onset of instability. 

Clearly, the stationary solutions of Eqs. (\ref{F1})(\ref{F2})(\ref{F3}) correspond to the isothermal gaseous spheres studied in the previous sections. It should be recalled that the condition of mechanical equilibrium corresponds to the balance between the pressure force and the gravitational force. For an isothermal gas, this yields
\begin{equation}
T{d\rho\over dr}+\rho{d\Phi\over dr}=0.
\label{F4}
\end{equation}
We now consider a small perturbation around a stationary solution and write
\begin{equation}
{\bf v}=\delta {\bf v}({\bf r},t),\qquad \rho=\overline{\rho}+\delta \rho({\bf r},t),
\label{F5}
\end{equation}
\begin{equation}
p=\overline{p}+\delta p({\bf r},t),\qquad \Phi=\overline{\Phi}+\delta \Phi({\bf r},t),
\label{F6}
\end{equation}
where the bar refers to the stationary solution (in the following we shall drop the bar). The linearized equations for the perturbations are  
\begin{equation}
\rho{\partial {\delta{\bf v}}\over \partial t}=-{kT\over m}\nabla\delta\rho-\delta\rho\nabla\Phi-\rho\nabla\delta\Phi,
\label{d6}
\end{equation}
\begin{equation}
{\partial \delta \rho\over \partial t}+\nabla (\rho {\delta{\bf v}})=0,
\label{d7}
\end{equation}
\begin{equation}
\Delta\delta\Phi=4\pi G\delta\rho.
\label{d8}
\end{equation}
Since we work with exact stationary solutions of the fluid equations,
that are inhomogeneous, we do not have to advocate the Jeans swindle
(Binney \& Tremaine 1987). In this sense, our approach is more
rigorous than the classical treatment of Jeans starting from an
infinite and homogeneous medium \footnote{It should be recalled,
however, that the Jeans procedure is justified in a cosmological
context if we take into account the expansion of the universe.}.

In the following, we restrict ourselves to radial perturbations. It is known that non-radial perturbations do not lead to new instabilities. Writing the time dependance of the perturbation in the form $\delta v\sim e^{\lambda t}$, $\delta\rho\sim e^{\lambda t}$,..., the equations of the problem become
\begin{equation}
\lambda\rho\delta v=-{kT\over m}{d\delta\rho\over dr}-\delta\rho{d\Phi\over dr}-\rho{d\delta\Phi\over dr},
\label{d9}
\end{equation}
\begin{equation}
\lambda \delta \rho+ {1\over r^{2}}{d\over dr}(\rho r^{2}\delta{v})=0,
\label{d10}
\end{equation}
\begin{equation}
 {1\over r^{2}}{d\over dr}\biggl ( r^{2}{d\delta\Phi\over dr}\biggr )=4\pi G\delta\rho.
\label{d11}
\end{equation}
Introducing the notation (\ref{s3}) and using the Gauss theorem (\ref{s7}) we see that the Poisson equation (\ref{d11}) is automatically satisfied. The continuity equation (\ref{d10}) becomes
\begin{equation}
{\lambda\over 4\pi r^{2}} {dq\over dr}+ {1\over r^{2}}{d\over dr}(\rho r^{2}\delta{v})=0.
\label{d12}
\end{equation}
This equation is readily integrated. Using the boundary condition $q(0)=0$, we get
\begin{equation}
\delta v=-{\lambda\over 4\pi \rho r^{2}}q.
\label{d13}
\end{equation}
Substituting this result back into Eq. (\ref{d9}), we obtain
\begin{equation}
 {\lambda^{2}\over 4\pi r^{2}}q={kT\over m}{d\over dr}\biggl ({1\over 4\pi r^{2}}{dq\over dr}\biggr )+{1\over 4\pi r^{2}}{dq\over dr}{d\Phi\over dr}+{G\rho\over r^{2}}q.
\label{d14}
\end{equation}
Using the condition of mechanical equilibrium (\ref{F4}), the foregoing equation can be rewritten
\begin{equation}
{k\over m}{d\over dr}\biggl ({1\over 4\pi \rho r^{2}}{dq\over dr}\biggr )+{Gq\over T r^{2}}={\lambda^{2}\over 4\pi \rho Tr^{2}}q.
\label{d15}
\end{equation}
This is similar to the eigenvalue equation (\ref{s10}) associated with the second variations of the free energy. In particular they coincide for marginal stability $\lambda=0$. Therefore, dynamical and thermodynamical instability occur at the same point in the series of equilibrium. This was not obvious {\it a priori} since the Navier-Stokes equations without viscosity conserve the free energy.  Taking into account a finite viscosity, we obtain instead of Eq. (\ref{d15}),
\begin{eqnarray}
{k\over m}{d\over dr}\biggl ({1\over 4\pi \rho r^{2}}{dq\over dr}\biggr )+{Gq\over T r^{2}}={\lambda^{2}\over 4\pi \rho Tr^{2}}q
-{1\over\rho T}\bigl (\zeta+{4\eta\over 3}\bigr ){d\over dr}\biggl \lbrack {1\over r^{2}}{d\over dr}\biggl ({\lambda q\over 4\pi \rho}\biggr )\biggr \rbrack. 
\label{d15bis}
\end{eqnarray}
For $\lambda=0$ the viscous term cancels out. Therefore, viscosity does not change the onset of instability, nor does it alter the profile of the perturbation that triggers the instability.

\begin{figure}[htbp]
\centerline{
\psfig{figure=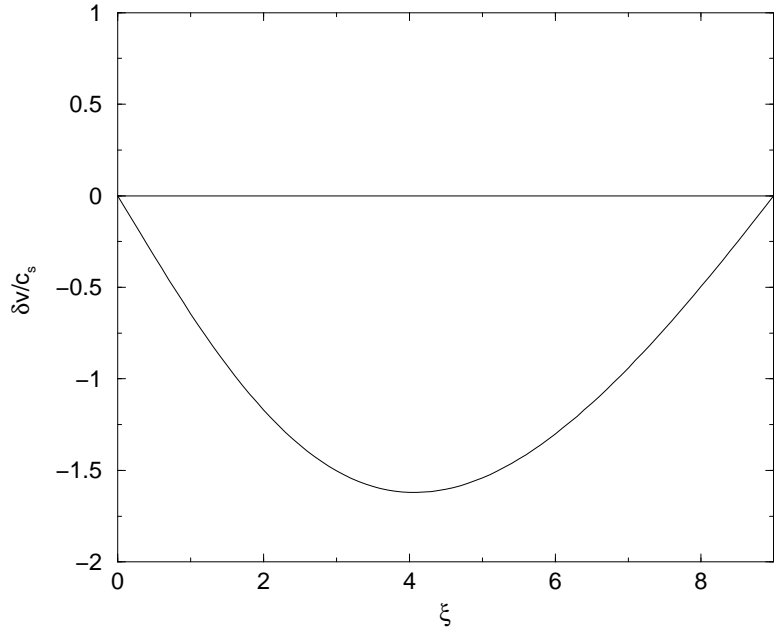,angle=0,height=8.5cm}}
\caption{Velocity profile for the first mode of instability corresponding to $\alpha_{1}=8.99$.}
\label{deltav1}
\end{figure}

\begin{figure}[htbp]
\centerline{
\psfig{figure=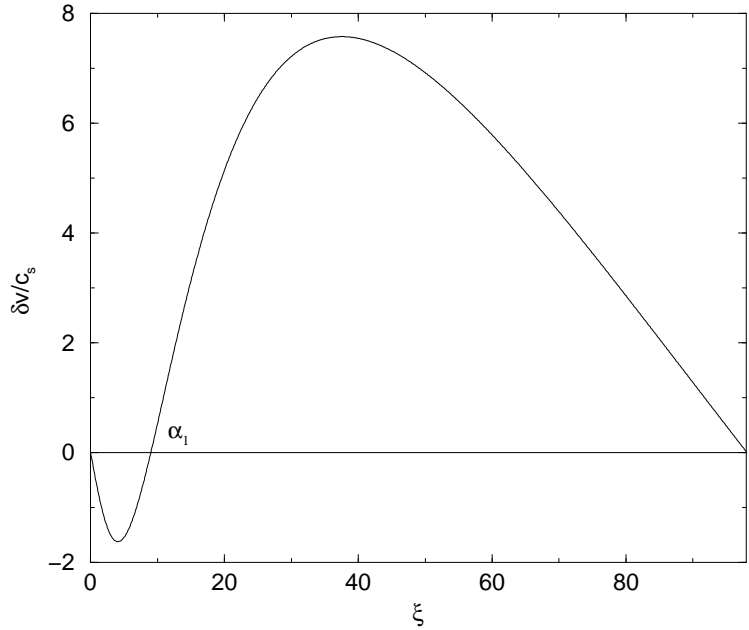,angle=0,height=8.5cm}}
\caption{Velocity profile for the second mode of instability corresponding to $\alpha_{2}=98.0$. We clearly see the separation between a core $(\delta v<0)$ and a halo ($\delta v>0$). }
\label{deltav2}
\end{figure}

The profiles of the density perturbation have already been discussed
in section \ref{sec_profile}. The profiles of the velocity
perturbation at the critical points are given by Eq. (\ref{d13}). In
order to work with dimensionless variables, we introduce the speed of
sound
\begin{equation}
c_{s}^{2}={dp\over d\rho}={k\over m}T={1\over\beta}.
\label{d16}
\end{equation}
Note that $c_{s}$ can be written $c_{s}={1\over 3}L_{K}/t_{dyn}$ where $t_{dyn}=(4\pi G\rho_{0})^{-1/2}$ is a time scale comparable with the dynamical time (Binney \& Tremaine 1987). Then
\begin{equation}
{\delta v\over c_{s}}=-{\lambda'\over 4\pi\xi^{2}}e^{\psi}F(\xi),
\label{d17}
\end{equation}
where we have set $\lambda'=\lambda t_{dyn}$. Using Eqs. (\ref{s17})-(\ref{uv}), we find
\begin{equation}
{\delta v\over c_{s}}=-{\lambda'\over 4\pi}c_{1}\psi'e^{\psi}(u-1).
\label{d18}
\end{equation}
We assume that we are just at the onset of the instability  ($\lambda'=0^{+}$) so that Eq. (\ref{d18}) is applicable with $\lambda'>0$ (the velocity profile at marginal stability, i.e. $\lambda'=0$, is simply $\delta v=0$). Since $\psi'(0)=0$, the velocity always vanishes at the center of the sphere $\delta v(\xi=0)=0$. The other zero(s) are determined by the condition $u(\xi_{i})=1$. Since the critical points of instability  are also characterized by $u(\alpha)=1$, see section \ref{sec_marginal}, we deduce that the  velocity perturbation always vanishes at the wall: $\delta v(\xi=\alpha)=0$. This result is to be expected since the mass does not leave the sphere. The other zeros are determined by the intersections between the spiral in the $(u,v)$ plane and the line $u=1$ (see Fig. \ref{uvosci}). They correspond precisely to the values of $\alpha$ at which a new mode of instability occurs, i.e. $\xi_{i}=\alpha_{i}$. It is therefore straightforward to determine the number of zeros in the velocity profile. For the first instability there are only two zeros at $\xi=0$ and $\xi=\alpha_{1}$. For the second instability there are three zeros at $\xi=0$, $\xi=\alpha_{1}$ and $\xi=\alpha_{2}$. For the $n^{th}$ instability there are $n+1$ zeros $\xi=0$, $\xi=\alpha_{1}$, $\xi=\alpha_{2}$,..., $\xi=\alpha_{n}$. The exact velocity profiles given by Eq. (\ref{d18}) are displayed on Figs. \ref{deltav1}-\ref{deltav2} for $n=1,2$. They are in agreement with the numerical results of Semelin {\it et al.} (2001). For high order modes of instability, the zeros follow asymptotically the geometric progression (\ref{g4}). Returning to dimensional variables, the zeros of the $n^{th}$ mode of instability are located at $r=0$ and $r_{i}=(\alpha_{i}/\alpha_{n})R\ (i=1,...,n)$.

\section{Fragmentation and fractal structure of an isothermal self-gravitating gas}
\label{sec_frag}

We shall now express the previous results in a more physical form and make some speculations about the fragmentation and the fractal structure of an isothermal self-gravitating gas.

\subsection{The King's length}
\label{sec_king}

Introducing the King's radius (Binney \& Tremaine 1987)
\begin{equation}
L_{K}=\biggl ({9\over 4\pi G\beta\rho_{0}}\biggr )^{1/2},
\label{g9}
\end{equation}
the parameter $\alpha$ defined in section \ref{sec_diagram} can be rewritten
\begin{equation}
\alpha=3{R\over L_{K}}.
\label{g10}
\end{equation}
The King's radius gives a good estimate of the core radius of an isothermal sphere. In the following, we shall consider that the core radius $L_{K}$ is fixed and that the domain size $R$ is increased. Clearly, this amounts to increasing the parameter $\alpha$ along the spiral. From the study of section \ref{sec_marginal}, we find that the onset of instability corresponds to 
\begin{equation}
R>{8.99\over 3}L_{K}.
\label{g11}
\end{equation}
For larger radii there are still critical points of free energy but they are more and more unstable. According to Eq. (\ref{g4}), a new mode of instability occurs at radii that follow asymptotically the geometric progression
\begin{equation}
R_{n}\sim [10.74...]^{n} L_{K}.
\label{g12}
\end{equation}
This result is clearly related to that of Semelin {\it et al.} (1999) who first noticed the appearance of a hierarchy of scales connected to the instability of isothermal gaseous spheres. However, the connection with their work is not straightforward. Semelin  {\it et al.} (1999) approximate the regular solutions of Eq. (\ref{Mfr})  by the analytical profile (\ref{singular}) and introduce an { arbitrary} short cut-off distance $\delta$ to avoid the central singularity. By contrast, there is no cut-off at small scales in our theory since we consider regular spheres that are exact solutions of the Emden equation (although we do not have analytical expressions for them). In this sense, there are no approximation nor indetermination in our theory. In fact, an ``effective'' cut-off  is played by the King's length (or core radius) but it is not arbitrary and depends on physical parameters,  the central density $\rho_{0}$ and the temperature $T$. For $r\gg L_{K}$, the density profile behaves like $r^{-2}$ so that regular gaseous spheres can indeed be approximated by a $r^{-2}$ profile plus a cut-off $\delta$ at small scales as done by Semelin {\it et al.} (1999). Our approach suggests to take $\delta$ as the King's length. We have therefore given a new derivation of their result which is considerably simpler and does not introduce sophisticated renormalization group technics. It also avoids the introduction of an arbitrary cut-off. This provides, hopefully, an easier interpretation of their intriguing results. 

According to section \ref{sec_profile}, if we increase the domain size along the series of equilibrium (i.e. maintaining a fixed core radius) smaller and smaller regions become unstable.  The modes that trigger these instabilities present numerous oscillations whose zeros also follow a geometric progression.  These oscillations can be regarded as sort of `germs' leading to the formation of `clumps' preceding the fragmentation of the isothermal gas. It is expected that these `clumps' will evolve by achieving higher and higher density contrasts, and finally fragment in turn into substructures. This yields a hierarchy of structures fitting one into each other in a self-similar way. This picture is given further support by the fact that {\it both}  the domain sizes that induce instability and the zeros of the perturbation in a given domain follow a geometric progression with the same ratio (see Eqs. (\ref{g12})(\ref{P5})). This property may  explain in a quite natural fashion the fractal structure that a self-gravitating medium (e.g., the interstellar medium, the large scale structures of the universe,...) can build under certain conditions. 

The hierarchy of scales that appears in an isothermal gas is due intrinsically to the oscillating nature of the equilibrium phase diagram (see Figs. \ref{gravotherme},\ref{isocollapse},\ref{uvcrit}). This curious behavior has been known for a long time in the context of stellar structure (Chandrasekhar 1942) but it had not been apparently related to the geometric progression of scales inducing instability.

\subsection{The Jeans length}
\label{sec_jeans}

We can also introduce another length scale defined with the average density $\overline{\rho}={3M\over 4\pi R^{3}}$ of the system instead of the central density. This is the Jeans length (Binney \& Tremaine 1987)
\begin{equation}
L_{J}=\biggl ({9\over 4\pi G\beta \overline{\rho}}\biggr )^{1/2}.
\label{g13}
\end{equation}
The reduced temperature $\eta$ can be expressed in terms of the Jeans 
length as (de Vega \& Sanchez 2001a)
\begin{equation}
\eta={\beta GM\over R}={3}\biggl ({R\over L_{J}}\biggr )^{2}.
\label{g14}
\end{equation}
In the following, we shall consider that the Jeans length $L_{J}$ is fixed and that the domain size $R$ is increased. Clearly, this amounts to increasing the parameter $\eta$ along the vertical axis of Fig. \ref{isocollapse}. Using the results of section \ref{sec_katz}, we find that the gaseous sphere becomes unstable when 
\begin{equation}
R>\sqrt{2.52...\over 3} L_{J}.
\label{g15}
\end{equation}
For larger radii the system will collapse because there are {\it no} critical point of free energy (i.e. hydrostatic equilibrium). In that case there are clearly no secondary instabilities.  Indeed, when the condition (\ref{g15}) is realized we leave the spiral and loose the geometric hierarchy of scales associated with its turning points. Since the mode of instability at the critical point does not oscillate (see section \ref{sec_profile}), we expect that the system will collapse {\it without} fragmenting, probably like in the study of Penston (1969).

\subsection{Fragmentation or not fragmentation?}
\label{sec_interpret}

The fragmentation of the isothermal self-gravitating gas and the
fractal structure that it can generate were previously discussed by de
Vega {\it et al.} (1996a, 1996b, 1998) and we here improve and
complete their physical arguments. In particular, the analytical
calculation of the modes of instablity (without approximation), the
geometric progression of {\it both} the domain sizes and the `clumps'
sizes, and the precise physical picture that emerges from these
results are new. In addition, we give arguments explaining why the
instability of a self-gravitating gas does not always lead to a
fractal structure (indeed globular clusters or elliptical galaxies,
for example, do not show a hierarchical structure). In our model, the
selection of the regime is related to how the size of the domain
compares with the Jeans length or the King's length. If $R>0.916...L_{J}$,
we expect a collapse without fragmentation but if $R>2.996...L_{K}$ (with
a sufficiently large ratio) the medium is expected to fragment into 
several `clumps'.

\begin{figure}[htbp]
\centerline{
\psfig{figure=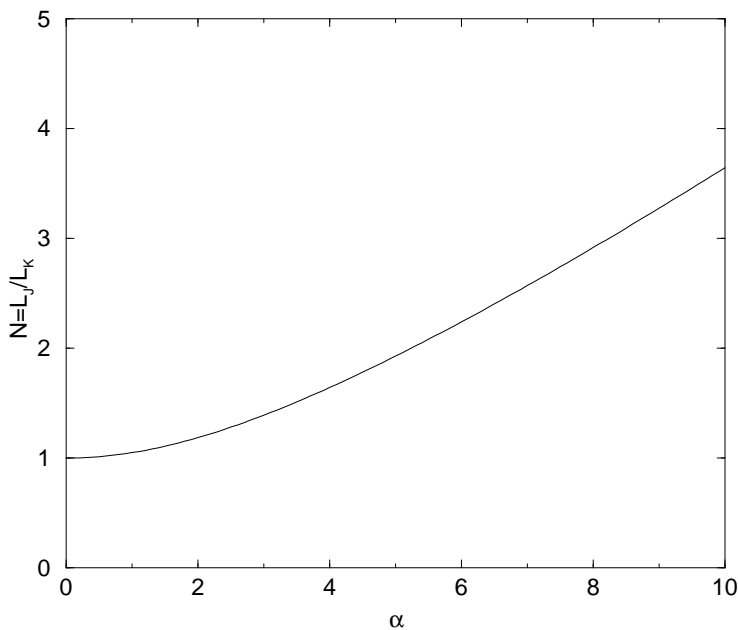,angle=0,height=8.5cm}}
\caption{Evolution of the dimensionless number ${\cal N}$ as a function of the scaled radius $\alpha$. This number quantifies the inhomogeneity of an isothermal self-gravitating gas.}
\label{N}
\end{figure}

The distinction between the Jeans scale and the King's scale is
therefore essential in the present context and had not been noticed
previously. In particular it is not the Jeans lengths but the King's
lengths that follow a geometric progression [see formula (\ref{g12})]
contrary to the claim of Semelin {\it et al.} (1999). These two
scales only coincide for a perfect gas with no self-gravity that is
homogeneous. Indeed, for $\alpha\rightarrow 0$, we have $L_{J}\sim
L_{K}\sim {3R\over\alpha}\rightarrow +\infty$. By contrast, for
$\alpha\rightarrow +\infty$ (corresponding to the singular sphere) we
get $L_{K}\sim {3R\over\alpha}\rightarrow 0$ and $L_{J}\rightarrow
\sqrt{3\over 2}R$ (we have used the asymptotic expansions of section
\ref{sec_diagram}). To appreciate the difference between the two
length scales, it is convenient to introduce a dimensionless number
\begin{equation}
{\cal N}={L_{J}\over L_{K}}.
\label{g16}
\end{equation}
From the above formulae, we have the equivalent expressions
\begin{equation}
{\cal N}=\biggl ({\rho_{0}\over \overline{\rho}}\biggr )^{1/2}={\alpha\over \sqrt{3\eta}}=\biggl ({\alpha\over 3\psi'(\alpha)}\biggr )^{1/2}.
\label{g17}
\end{equation}
For $\alpha\rightarrow 0$ (perfect gas), ${\cal N}\sim 1$ and for $\alpha\rightarrow +\infty$ (singular sphere), ${\cal N}\sim {\alpha\over \sqrt{6}}\rightarrow +\infty$. Therefore the number ${\cal N}$ quantifies the inhomogeneity of the self-gravitating gas. The evolution of ${\cal N}(\alpha)$ along the spiral is ploted on Fig. \ref{N}. From Eq. (\ref{g17}), ${\cal N}$ is just the ratio between the central density and the average density to the power $1/2$.

\section{Conclusion}
\label{sec_conclusion}

In the first part of the paper, we have investigated the stability of isothermal spheres in the canonical ensemble using a thermodynamical approach. This is a natural extension of the work of Padmanabhan (1989) in the microcanonical ensemble. As is well known, there is no hydrostatic equilibrium for an isothermal gas below a critical temperature  $kT_{c}={GmM\over 2.52 R}$ (Lynden-Bell \& Wood 1968). This corresponds to a situation in which the domain size becomes comparable with the Jeans length, i.e. $R> 0.916...L_{J}$. In that case, we expect the system to collapse. By analyzing the second variations of free energy, we have found that the point of minimum temperature is precisely the point at which the free energy ceases to be a local maximum and becomes a saddle point in the series of equilibrium. This vindicates the theorem of Katz (1978) and provides in addition the form of the perturbation that induces instability. Contrary to the microcanonical ensemble, it has not a ``core-halo'' structure, nor multiple oscillations. We expect therefore the system to collapse without fragmenting.  On the other hand, if we start from isothermal gaseous spheres with high density contrasts ${\cal R}>32.2$ (this corresponds to $R>2.996...L_{K}$), such spheres are unstable saddle points and probably fragment into  small `clumps' (although they would be said to be stable by a naive Jeans argument since $R< 0.916...L_{J}$). The number of clumps, presumably associated with the number of zeros in the mode of instability, increases with the density contrast or, equivalently, as the domain size becomes larger and larger with respect to the King's radius. The distinction between the Jeans length and the King's length  is essential in the present context and had not been noticed previously. The Jeans length is related to the parameter $\eta$ determining  whether an  hydrostatic equilibrium state exists or not (for an isothermal sphere) while the King's length is related to the parameter $\alpha$ parametrizing the series of equilibrium. They are very different parameters.  

In the second part of the paper, we have extended Jeans instability criterion to the case of inhomogeneous bounded gaseous spheres described by Navier-Stokes equations. This treatment avoids the well-known difficulties associated with an infinite and homogeneous medium (Jeans swindle). Quite remarkably, we have provided an elegant analytical solution of this more general problem without making any approximation.

Now, the question that naturally emerges is what happens to a
gravitationally unstable gaseous configuration. The collapse of
isothermal gas spheres has been investigated by Penston (1969) using
Navier-Stokes equations (without viscosity). He found an analytical
solution for which the collapse is self-similar and develops a finite
time singularity (i.e., the central density becomes infinite in a
finite time). This only {\it partially} answers the above
question. Indeed, his solution is just a particular solution of the
fluid equations and other solutions (e.g., fragmentation) may be
possible. In particular, it is not clear whether his solution is
stable with respect to non spherically symmetric perturbations. In
addition, Penston (1969) works in an infinite domain for which the
isothermal spheres have infinite mass and are always
unstable. Therefore, he could not evidence a phase transition between
an equilibrium ``gaseous'' state and a ``collapsed'' state, depending
on the value of the temperature.

The thermodynamics of self-gravitating systems is so intriguing that
we have been tempted to explore this problem anew from a dynamical
point of view (Chavanis {\it et al.} 2001). We have introduced a
numerical algorithm in the form of a relaxation equation (Chavanis
1996, 2001a; Chavanis {\it et al.} 1996) constructed so as to increase
monotonically entropy at fixed energy in the microcanonical ensemble
and free energy at fixed temperature in the canonical ensemble. With
this numerical algorithm, we have covered the whole
bifurcation diagram in parameter space and we have checked, by an
independant method, the stability limits of Katz (1978). We have
verified that the density profile that induces instability at the
critical point has a ``core-halo'' structure in the microcanonical
ensemble but not in the canonical ensemble, in agreement with the
study of Padmanabhan (1989) and the present study. We have also
checked that the number of oscillations in the perturbation profile
increases when we start from unstable isothermal spheres with high
density contrasts. With this algorithm, we have explored the ``bassin
of attraction'' of the stable isothermal spheres. Since they only are
{\it local} maxima of entropy or free energy, an initial condition can
either relax towards a metastable equilibrium state or collapse,
depending on its {\it topology}. Hence, the control parameter
$\Lambda$ or $\eta$ is not sufficient to characterize completely the
final state of the system even in the case when an equilibrium
exist. This depends whether the initial condition lies in the ``bassin
of attraction'' of the equilibrium solution or not. The complete
characterization of this bassin of attraction is difficult but we have
given some examples. With our numerical algorithm, it will be possible
to include rotation in order to compute entropy maxima that are not
spherically symmetric and determine the influence of angular momentum
conservation on the collapse of a rotating self-gravitating system
(Chavanis \& Rosier, in preparation). Some interesting works have been
done in that direction (Lagoute \& Longaretti 1996, Laliena 1999,
Fliegans \& Gross 2001 and Lynden-Bell 2000) but the complete
description of rotating isothermal gaseous spheres is still in its
infancy stage.

It is also of first interest to test numerically the speculations that
we have made concerning the development of the instability, i.e. if
the system will collapse as a whole or if it will fragment and break
into substructures. These two regimes are clearly separated, at least
in the theoretical model that we have considered here (Antonov
model). We have suggested that if we start from a solution of the
Emden equation with a high density contrast, it should break in a
series of `clumps' associated with the oscillations of the density
perturbation that we have calculated in the linear regime. Obviously,
if we want to follow the nonlinear evolution correctly, we must relax
the hypothesis of spherical symmetry.

In this paper, we have assumed that the particles can be treated by
classical mechanics and that they are point-like. The extension of our
work to general relativity (for, e.g., neutron stars) is almost
straightforward (Chavanis 2001b). We find that there is no hydrostatic
equilibrium if the system size is smaller than a multiple of the
Schwarzschild radius. We also show analytically that the point of
smallest radius corresponds precisely to the point at which the series
of equilibrium becomes unstable. Secondary modes of instability are
found; they follow a geometric progression like in the Newtonian case
but with a different ratio. When the speed of light goes to $+\infty$,
we recover the results of the present study. It is seen that
relativistic effects {\it favour} gravitational instability,
i.e. instability occurs sooner than in the Newtonian limit. On the
other hand, the case of self-gravitating fermions enclosed within a
box has been considered by Chavanis \& Sommeria (1998). Their study
was formulated in the context of the ``violent relaxation'' of
collisionless stellar systems introduced by Lynden-Bell (1967) but the
results also apply to quantum particles such as massive neutrinos in
Dark Matter models (Ingrosso {\it et al.} 1992) or degenerate
electrons in White Dwarf stars (Chandrasekhar 1959). When degeneracy
is accounted for, there exists a {\it global} entropy maximum for all values
of energy (Chavanis \& Sommeria 1998, Robert 1998). It has a
``core-halo'' structure with a degenerate core and a dilute
atmosphere. Depending on the degeneracy parameter, there can be a
``gravothermal catastrophe'' at $E=-0.335GM^{2}/R$ but the core ceases
to shrink when it becomes degenerate. Considering a classical gas with
a short distance cut-off $a$ essentially leads to the same results
(see, e.g., Padmanabhan 1990, Follana \& Laliena 2000).

For astrophysical purposes, it is still a matter of debate to decide whether collisionless stellar systems like elliptical galaxies are degenerate (in the sense of Lynden-Bell) or not. Since degeneracy can stabilize the system without changing its overall structure at large radii, we have suggested that degeneracy could play a role in galactic nuclei  (Chavanis \& Sommeria 1998). The recent simulations of Leeuwin and Athanassoula (2000) and the theoretical model of Stiavelli (1998) seem to go in that direction: if the nucleus of elliptical galaxies contains a (primordial) black hole, degeneracy must be taken into account and can explain the cusps observed in the center of galaxies. This form of degeneracy is also relevant for massive neutrinos in Dark Matter models where it competes with quantum degeneracy (Kull {\it et al.} 1996).

All together, these results suggest that the statistical mechanics and the thermodynamics of self-gravitating systems can have quite relevant astrophysical applications at different scales of the universe:  stars (Chandraskhar 1959), interstellar medium (de Vega {\it et al.} 1996a, 1996b),  globular clusters (Lynden-Bell \& Wood 1968), elliptical galaxies (Lynden-Bell 1967, Hjorth \& Madsen 1993, Chavanis \& Sommeria 1998), dark matter (Ingrosso {\it et al.} 1992, Kull {\it et al.} 1996), cosmology (Saslaw \& Hamilton 1984, de Vega {\it et al.} 1998)...The same ideas of statistical mechanics have been introduced in two-dimensional turbulence to explain the formation and maintenance of coherent vortices, like Jupiter's Great Red Spot, which are common features of large-scale geophysical and astrophysical flows (see Bouchet \& Sommeria 2000, Bouchet {\it et al.} 2001 and references therein). The analogy between the statistical mechanics of two-dimensional vortices and stellar systems has been described extensively by Chavanis (1996, 1998b, 2001c). This analogy concerns not only the equilibrium state (the formation of large-scale structures) but also the relaxation towards equilibrium (Chavanis {\it et al.} 1996, Chavanis 2001d) and the statistics of fluctuations (Chavanis \& Sire 2000). These ideas may also have applications in the context of planet formation where large-scale vortices present in the solar nebula  could efficiently trap  dust particles to form the planetesimals and the planets (Barge \& Sommeria 1995, Tanga {\it et al.} 1996, Bracco {\it et al.} 1999, Chavanis 2000). Accordingly, the statistical mechanics of these nonlinear media seems to be able to account for the fascinating process of self-organization in nature.

\section{Acknowledgments}
\label{sec_ack}

I acknowledge interesting discussions with L. Athanassoula, F. Combes, O. Fliegans, J. Hjorth, H. Kandrup, V. Laliena, F. Leeuwin, D. Lynden-Bell, T. Padmanabhan, R. Robert, B. Semelin and J. Sommeria  on these exciting problems of thermodynamics. This work was initiated during my stay at the Institute for Theoretical Physics, Santa Barbara, during the program on Hydrodynamical and Astrophysical Turbulence (February-June 2000). This research was supported in part by the National Science Foundation under Grant No. PHY94-07194.


\begin{thebibliography}{10}


\bibitem{antonov}  {\small Antonov V.A., 1962, Vest. Leningr. Gos. Univ. {7}, 135}

\bibitem{barge}  {\small  Barge P., Sommeria J., 1995, ``Did planet formation begin inside persistent gaseous vortices?'' A\&A  {295}, 
L1-L4}

\bibitem{bouchet1}  {\small  Bouchet F., Sommeria J., 2000, ``Emergence of intense jets and Jupiter Great Red Spot as maximum entropy structures'', submitted to J. Fluid. Mech. [physics/0003079]}

\bibitem{bouchet2}  {\small  Bouchet F., Sommeria J., Chavanis P.H.,``Statistical mechanics of Jupiter's Great Red Spot in the shallow water model'' in preparation}

\bibitem{bt}  {\small  Binney J., Tremaine S., 1987,
Galactic Dynamics (Princeton Series in Astrophysics)} 

\bibitem{bracco}  {\small Bracco A., Chavanis P.H.,  Provenzale A., Spiegel E.A., 1999, ``Particle aggregation in a turbulent Keplerian flow'', Phys. Fluids {11}, 2280}

\bibitem{cerruti}  {\small Cerruti-Sola M., Cipriani P., Pettini M., 2001, ``On the clustering phase transition in self-gravitating $N$-body systems'', submitted to MNRAS [astro-ph/0105380]}

\bibitem{chandra}  {\small  Chandrasekhar S., 1942, 
An Introduction to the Theory of Stellar Structure (Dover)} 

\bibitem{cthese}  {\small  Chavanis P.H., 1996, Contribution \`a la m\'ecanique statistique des tourbillons bidimensionnels. Analogie avec la relaxation violente des syst\`emes stellaires. Th\`ese de doctorat. Ecole Normale Sup\'erieure de Lyon}

\bibitem{quasi}  {\small  Chavanis P.H., 1998a,  ``On the coarse-grained evolution of collisionless stellar systems'', MNRAS  {300}, 981}

\bibitem{cfloride}  {\small Chavanis P.H., 1998b, ``From Jupiter's Great Red Spot to the structure 
of galaxies: statistical mechanics of two-dimensional 
vortices and stellar systems'',  Annals N.Y. Acad. Sci. {867}, 120}

\bibitem{cplanetes}  {\small  Chavanis P.H., 2000, ``Trapping of dust by coherent vortices in the solar nebula'', A\&A  {356}, 1089}

\bibitem{dubrovnik}  {\small  Chavanis P.H., 2001a, Statistical mechanics of violent relaxation in stellar systems. In: Proceedings of the Conference on Multiscale Problems in Science and Technology (Springer)}

\bibitem{chavRG}  {\small Chavanis P.H., 2001b, ``Gravitational instability of finite isothermal spheres in general relativity'', in preparation}

\bibitem{japon}  {\small Chavanis P.H., 2001c, On the analogy between two-dimensional vortices and stellar systems. In: Proceedings of the IUTAM Symposium on Geometry and Statistics of Turbulence, T. Kambe, T. Nakano and T. Miyauchi Eds. (Kluwer Academic Publishers)}

\bibitem{kinetic}  {\small Chavanis P.H., 2001d, ``Kinetic theory of point vortices: diffusion coefficient and systematic drift'', to appear in Phys. Rev. E [cond-mat/0107219]}

\bibitem{cr}  {\small  Chavanis P.H., Rosier C., Sire C. 2001, ``Thermodynamics of self-gravitating systems'', submitted to Phys. Rev. E [cond-mat/0107345] }

\bibitem{csire1}  {\small Chavanis P.H., Sire C. 2000, ``Statistics of velocity fluctuations arising from a random distribution of point vortices: the speed of fluctuations and the diffusion coefficient'', Phys. Rev. E {62}, 490}

\bibitem{cs}  {\small Chavanis P.H., Sommeria J., 1998, ``Degenerate equilibrium states of collisionless 
stellar systems'',  MNRAS  {296}, 569}

\bibitem{csr}  {\small  Chavanis P.H., Sommeria J., Robert R., 1996, ``Statistical mechanics of two-dimensional 
vortices and collisionless stellar systems'' , Astrophys.
J. {471}, 385}

\bibitem{cohn}  {\small  Cohn H., 1980, ``Late core collapse 
in star clusters and the gravothermal instability'', 
ApJ {242}, 765}

\bibitem{vega0}  {\small  de Vega H.J.,  Sanchez N., Combes F., 1996a,  ``Self-gravity as an explanation of the fractal structure of the interstellar medium'', Nature {383}, 56}

\bibitem{vega1}  {\small de Vega H.J., Sanchez N. , Combes F., 1996b, ``Fractal dimensions and scaling laws in the interstellar medium: a new field theory approach'', Phys. Rev. D {54}, 6008}

\bibitem{vega2}  {\small  de Vega H.J.,  Sanchez N.,  Combes F., 1998, ``The fractal structure of the universe : a new field theory approach'', ApJ {500}, 8} 

\bibitem{vega3}  {\small de Vega H.J.,  Sanchez N., 2001a,  ``Statistical mechanics of the self-gravitating gas: I. Thermodynamical limit and phase diagrams'', [astro-ph/0101568]} 

\bibitem{vega4}  {\small de Vega H.J., Sanchez N., 2001b, ``Statistical mechanics of the self-gravitating gas: II. Local physical magnitudes and fractal structure'', [astro-ph/0101567]} 

\bibitem{dubrulle2}  {\small  Dubrulle B.,  Graner F., 1994,  ``Titius-Bode laws in the solar system II. Build your own law from disk models'', A\&A  {282}, 269} 

\bibitem{fliegans}  {\small  Fliegans O.,  Gross D.H.E., 2001, ``Effect of angular momentum on equilibrium properties of a self-gravitating system'', submitted to Phys. Rev. E. [cond-mat/0102062]} 

\bibitem{fl}  {\small  Follana E.,  Laliena V., 2000, ``Thermodynamics of self-gravitating systems with softened potentials'', Phys. Rev. E  {61}, 6270}
 
\bibitem{dubrulle1}  {\small  Graner F. , Dubrulle B., 1994, ``Titius-Bode laws in the solar system II. Build your own law from disk models'', A\&A  {282}, 262} 

\bibitem{hjorth}  {\small Hjorth J., Madsen J., 1993,  ``Statistical Mechanics of Galaxies'', MNRAS  {265}, 237}

\bibitem{ruffini}  {\small Ingrosso G., Merafina M., Ruffini R. , 
Strafella F., 1992, ``System of self-gravitating semidegenerate
fermions with a cutoff of energy and angular momentum in their
distribution function'', A\&A {258}, 223 }

\bibitem{katz}  {\small  Katz J., 1978,``On the number of 
unstable modes of an equilibrium'',   MNRAS  
{183}, 765 }

\bibitem{kull}  {\small  Kull A.,  Treumann R.A. , B\"oringer H., 1996, 
 ``Violent relaxation of indistinguishable objects and neutrino hot dark matter in clusters of galaxies'',  ApJ Lett.   {466}, L1 }

\bibitem{longaretti}  {\small  Lagoute C.,  Longaretti P.Y., 1996,
``Rotating globular clusters. I. Onset of the gravothermal instability'', A\&A {308}, 441 }

\bibitem{laliena}  {\small  Laliena V., 1999, ``Effect of angular momentum conservation in the phase transition of collapsing systems'',   Phys. Rev. E
{59}, 4786 }

\bibitem{larson}  {\small  Larson R.B., 1970, ``A method for 
computing the evolution of star clusters'',  MNRAS 
{147}, 323 }

\bibitem{leeuwin}  {\small  Leeuwin F.,  Athanassoula E., 2000, ``Central cusp caused by a supermassive black hole in axisymmetric 
models of elliptical galaxies'', 
MNRAS  {417}, 79}

\bibitem{lb}  {\small  Lynden-Bell D., 1967, ``Statistical mechanics 
of violent relaxation in stellar systems'', MNRAS  {136}, 101}

\bibitem{lbrot}  {\small  Lynden-Bell D., 2000,  ``Rotation, Statistical Dynamics and Kinematics of Globular Clusters'', [astro-ph/0007116]}

\bibitem{lbe}  {\small  Lynden-Bell D. , Eggleton P.P., 1980, ``On the consequences of the gravothermal catastrophe'', 
MNRAS  {191}, 483 }

\bibitem{lblb}  {\small  Lynden-Bell D. , Lynden-Bell R.M., 1977, ``On the negative specific heat paradox'', 
MNRAS  {181}, 405 }

\bibitem{lbw}  {\small  Lynden-Bell D. , Wood R., 1968, ``The gravothermal catastrophe in isothermal spheres and 
the onset of red-giants structure for stellar systems'', 
MNRAS  {138}, 495 }

\bibitem{ogorodnikov}  {\small  Ogorodnikov K.F., 1965, Dynamics of stellar systems (Pergamon)}

\bibitem{pad2}  {\small  Padmanabhan T., 1989,  ``Antonov instability 
and the gravothermal catastrophe-revisited'', ApJ Supp.  {71}, 651 }

\bibitem{pad}  {\small Padmanabhan T., 1990, ``Statistical mechanics 
of gravitating systems'', Phys. Rep.  {188}, 285 }

\bibitem{penston}  {\small Penston M.V., 1969,  ``Dynamics of self-gravitating gaseous spheres III. Analytical results in the free fall or isothermal cases'',  MNRAS   {144}, 425 }

\bibitem{pfenniger1}  {\small  Pfenniger D.,  Combes F. , Martinet L., 1994, ``Is dark matter in spiral galaxies cold gas? I. Observational constraints and dynamical clues about galaxy evolution  '',  A\&A  {285}, 79 }

\bibitem{pfenniger2}  {\small  Pfenniger D. , Combes F., 1994, ``Is dark matter in spiral galaxies cold gas? II. Fractal models and star non-formation'', A\&A  {285}, 94}

\bibitem{robert}  {\small Robert R., 1998, ``On the gravitational collapse 
of stellar systems'', Class. Quantum Grav.  {15}, 3827}

\bibitem{saslaw}  {\small  Saslaw W.C., Hamilton A.J.S., 1984, ``Thermodynamics and galaxy clustering - Nonlinear theory of high order correlations'', ApJ {276}, 13}

\bibitem{semelin2}  {\small Semelin B., de Vega H.J.,  Sanchez N. , Combes F.,  1999,  ``Renormalization group flow and fragmentation in the self-gravitational thermal gas'', Phys. Rev. D {59}, 125021}

\bibitem{semelin1}  {\small  Semelin B., Sanchez N., de Vega H.J., 2001, ``Self-gravitating fluid dynamics, instabilities and solitons'', Phys. Rev. D {63}, 084005}

\bibitem{st}  {\small Shapiro S.L. , Teukolsky S.A., 1995, ``The collapse of dense star clusters to supermassive black holes: the origin of quasars and AGNs'',  ApJ {292}, L41}

\bibitem{sti}  {\small Stiavelli M., 1998,  ``Violent relaxation around a massive black hole'', ApJ Lett.  {495}, L91}

\bibitem{tanga}  {\small Tanga P.,  Babiano A. , Dubrulle B.,
 Provenzale A., 1996, ``Forming planetesimals 
in vortices'' , Icarus {121}, 158 }

\bibitem{yabushita}  {\small Yabushita S., 1968, ``Jeans's type gravitational instability of finite isothermal gas spheres'',  MNRAS  {140} 109}

\end{thebibliography}
\end{document}